\documentclass{llncs}
\pdfoutput=1
\usepackage{url}
\usepackage{graphicx}
\usepackage{amsfonts}
\usepackage{subfig}
\usepackage{amsmath}
\usepackage{algorithm2e}
\usepackage{multirow}
\newcommand{\condge}{\texttt{cond\_ge}}
\newcommand{\condle}{\texttt{cond\_le}}

\newcommand{\step}{\textit{{step}}}
\newcommand{\op}{\textit{op}}
\newcommand{\fa}{\textit{FA}}
\newcommand{\att}{\textit{att}}
\newcommand{\timestamp}{\textit{{ts}}}
\newcommand{\ab}{\mathbb{A}}
\newcommand{\bb}{\mathbb{B}}

\newcommand{\ws}{\textit{{ws}}}
\newcommand{\aux}{\textit{aux}}
\newcommand{\ct}{\textit{CT}}
\newcommand{\pk}{\textit{PK}}
\newcommand{\sk}{\textit{SK}}
\newcommand{\mk}{\textit{MK}}
\newcommand{\tk}{\textit{TK}}
\newcommand{\g}{\mathbb{G}}
\newcommand{\z}{\mathbb{Z}_p}
\newcommand{\e}{\mathcal{E}}
\newcommand{\enc}{\mathsf{Enc}}
\newcommand{\dec}{\mathsf{Dec}}
\newcommand{\trans}{\mathsf{Trans}}
\newcommand{\keygen}{\mathsf{KeyGen}}
\newcommand{\la}{\langle}
\newcommand{\ra}{\rangle}
\newcommand{\TS}{\textit{TS}}
\newcommand{\mtext}[1]{\text{{#1}}}

\newcommand{\map}{\texttt{Map}}
\newcommand{\filter}{\texttt{Filter}}
\newcommand{\join}{\texttt{Join}}

\newcommand{\gen}{\mathsf{Gen}}
\newcommand{\adv}{\mathsf{Adv}}

\begin{document}

\title{Streamforce: Outsourcing Access Control Enforcement for Stream Data to the Clouds}
\author{Tien Tuan Anh Dinh \and Anwitaman Datta}
\institute{School of Computer Engineering, Nanyang Technological University, Singapore
\\ \{ttadinh,anwitaman\}@ntu.edu.sg}
\maketitle

\begin{abstract}
As tremendous amount of data being generated everyday from human activity and from devices equipped
with sensing capabilities, cloud computing emerges as a scalable and cost-effective platform to
store and manage the data. While benefits of cloud computing are numerous, security
concerns arising when
data and computation are outsourced to a third party still hinder the complete movement to the
cloud.  In this paper, we focus on the problem of data privacy on the cloud, particularly on access
controls over stream data. The nature of stream data and the complexity of
sharing data make access control a more challenging issue than in traditional archival databases. We
present Streamforce --- a system allowing data
owners to securely outsource their data to the cloud. The owner specifies fine-grained policies
which are enforced by the cloud. The latter performs most of the heavy computations, while
learning nothing about the data content. To this end, we employ a number of encryption schemes, including
deterministic encryption, proxy-based attribute based encryption and sliding-window encryption. In
Streamforce, access control policies are modeled as secure continuous queries, which entails minimal
changes to existing stream processing engines, and allows for easy expression of a wide-range of
policies. In particular, Streamforce comes with a number of secure query operators
including Map, Filter, Join and Aggregate. Finally, we implement Streamforce over an open-source
stream processing engine (Esper) and evaluate its performance on a cloud platform. The results
demonstrate practical performance for many real-world applications, and although the security
overhead is visible, Streamforce is highly scalable.
\end{abstract}

\section{Introduction}
An enormous amount of data is being generated everyday, with sources ranging from traditional
enterprise systems to social applications. It becomes increasingly common to process such
data as they arrive in continuous streams. Examples range from high-frequency streams such as
generated from stock or network monitoring applications, to low-frequency streams originated from 
weather monitoring, social network or fitness monitoring~\footnote{\url{nikeplus.nike.com},
\url{fitbit.com}} applications. The variety and abundance of data, combined with the
potential of social interactivity, mash-up services and data sciences, has turned data
sharing into a new norm. A critical problem with sharing data is security, which concerns the
question of who gets access to which aspects of the data (fine-grained access control), and
under which context (data privacy). This paper studies the former question, which we believe to be
more challenging for stream data than for archival data because of three reasons. First, traditional
archival data systems enforce access control by pre-computing views, which is not possible with
stream data because of its infinite size.
Second, access control over stream is inherently data-driven (triggered by arrival of specific
data values) as opposed to user-driven with archival data, and it often involves temporal
constraints (sliding windows). Third, many of the sharing activities take place
in collaborative settings which entail a large number of users and even a larger number of policies. 

At the same time, cloud computing is driving a paradigm shift in the computing landscape. More
businesses and individual users are taking full advantage of the elastic, instantly available and
virtually unbounded computing resources provided by various vendors at competitive prices.  Many
enterprise systems are migrating their infrastructure to the cloud, while the convenience and
instant access to computing resource also spawns a plethora of small-to-medium size systems being
developed and deployed on the cloud. In the context of stream data sharing, cloud computing emerges
as an ideal platform for two reasons. First, data can be hosted and managed by a small number of cloud
providers with unlimited resources, which is important since data streams are of infinite sizes.
Second, data co-location makes it easy to share and to perform analytics. However, since data is
outsourced to untrusted third parties, enforcing access control on the cloud becomes
even more imperative and more challenging. 

In this paper, we present \emph{Streamforce} --- a fine-grained access control system for stream data over
untrusted clouds. Streamforce is designed with three goals. First, it supports specification and
enforcement of fine-grained access control policies. Second, data is outsourced to the cloud where
access control policies are enforced, with the latter learning nothing about the data content.
Third, the system is \emph{efficient}, in the sense that the cloud handles most of the expensive computations.
The last two goals require the cloud to be more active than being merely a storage facility.
To realize these goals, Streamforce uses a number of encryption schemes: deterministic encryption,
proxy-based attribute based encryption, and a sliding-window based encryption. While encryption is
necessary to protect data confidentiality against the cloud and against unauthorized access, we
believe that directly exposing encryption details to the system
entities (data owner, user and cloud) is not the ideal abstraction when it comes to access control.
Instead, Streamforce models access control policies using secure query operators: secure Map, Filter,
Join and Aggregate. These operators are at higher level and more human-friendly than raw encryption
keys. Enforcement at the cloud is the same as executing the secure queries. Since
existing stream processing engines are very efficient at executing continuous queries made from similar query
operators, they can be leveraged by the cloud without major changes. 

Streamforce occupies an unique position in the design space of outsourced access control. It
considers untrusted (semi-honest) clouds, which is different to~\cite{carminati10,dinh12}. Systems
such as Plutus~\cite{kallahalla03} and CryptDb~\cite{popa11} assume untrusted clouds, but they
support only coarse-grained policies over archival data. Recent systems utilizing attribute-based
encryption~\cite{goyal06,yu10} achieve more fine-grained access control on untrusted clouds, but
they do not support stream data.  Furthermore, the cloud is not fully utilized as it is used
mainly for storage and distribution. To the best of our knowledge, Streamforce is the first system
that allows secure, efficient outsourcing of fine-grained access control for stream data to untrusted
clouds. It is not catered for applications demanding high throughput, but it presents 
important first steps towards supporting them. Our contributions are summarized as follows: \begin{itemize}
\item We present a system and formal security model for outsourcing access control of stream data to
untrusted clouds. We discuss different security levels that different query operators can achieve. 
\item We present details and analyze security properties of different encryption schemes used for
fine-grained access control, including a new scheme supporting sliding window aggregation. 
\item We show how to use these encryption schemes to construct secure query operators: secure Map,
secure Filter, secure Join and secure Aggregate. 
\item We implement a prototype of Streamforce~\cite{streamforce} over Esper --- a high-performance stream processing engine. We then
benchmark it on Amazon EC2. The results indicate practical performance for many
applications. Although the cost of security is evident, we show that it can be compensated by the
system's high scalability.    
\end{itemize}
Next we present the system and security model, followed by the constructions of the
encryption schemes. We then describe how to construct secure query operators. Prototype
implementation and evaluation is presented in Section~\ref{sec:evaluation}. Related work follows in
Section~\ref{sec:relatedWork}, before we draw conclusion and discuss future work. 


\section{System and Security Model}
\label{sec:model}
\subsection{System Model}
\subsubsection{Overview.}
\begin{figure}
\centering
{\includegraphics[scale=0.45]{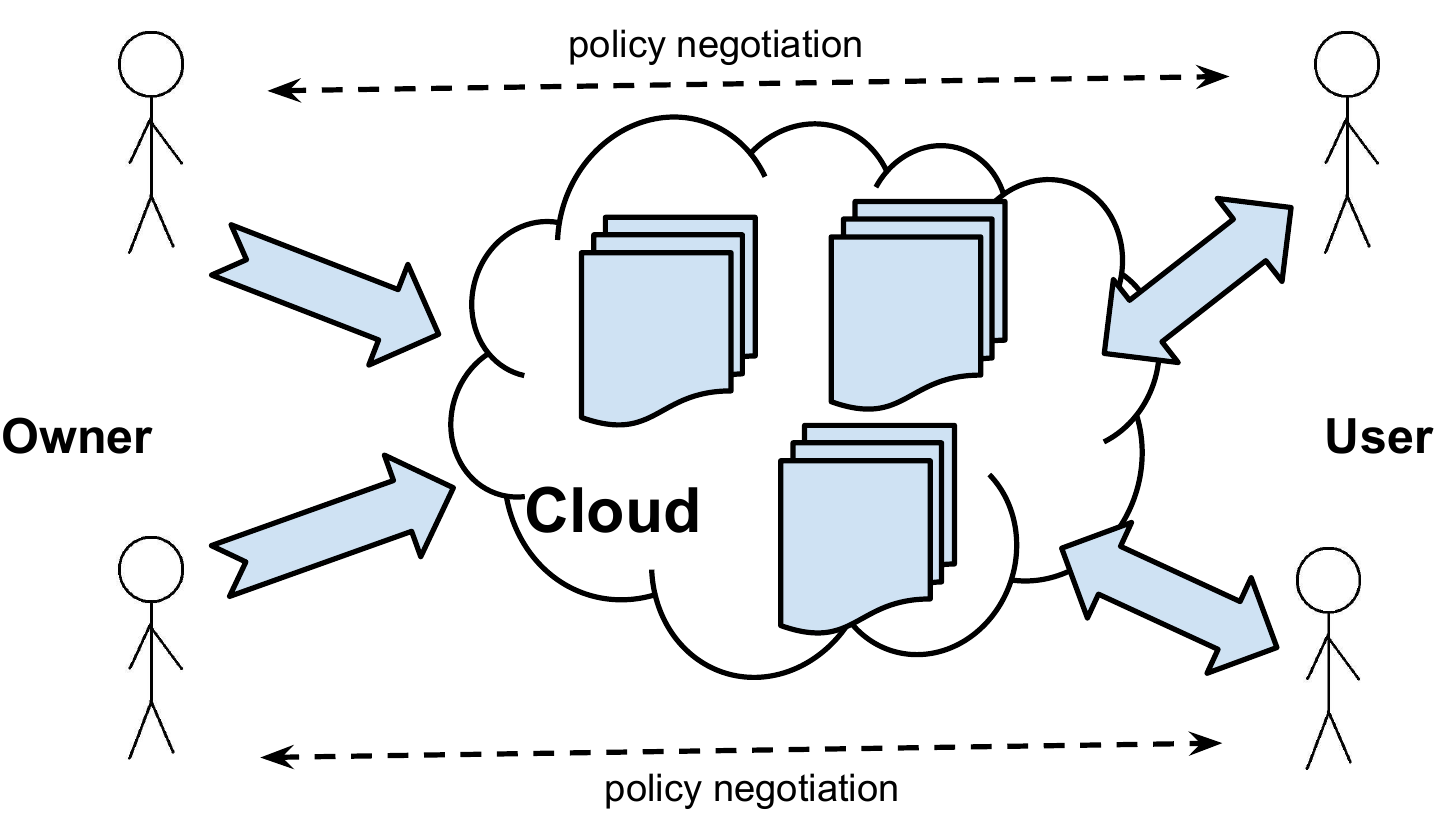}}
\caption{Overview of Streamforce's deployment}
\label{fig:systemModel}
\end{figure}
There are three types of entities: \emph{data owners} (or owners),  \emph{data users} (or users) and
a \emph{cloud}. Their interactions are illustrated in Fig.~\ref{fig:systemModel}: the owners
encrypt their data and relay them to the cloud, which performs transformation  and forwards the
results to the users for final decryption. We do not consider how the owner determines access
control policies, and we assume that the negotiation process (in which the owner grants policies to
the user) happens out-of-band. The system goals are three-folds: 
\begin{enumerate}
\item The owner is able to express flexible, fine-grained access control policies. 
\item The system ensures data confidentiality against untrusted cloud, and access control
against unauthorized users (as elaborated later).
\item Access control enforcement is done by the cloud. Decryptions at the user are
light-weight operations compared to the transformations at the cloud.   
\end{enumerate}

\subsubsection{Data Model.}
A data stream $S$ has the following schema:
\[
S = (\TS,A_1,A_2,..,A_n)
\]
where $\TS = \mathbb{N}$ is the timestamp, and all data attributes $A_i$ are of integer domains. A data
tuple at time $\timestamp$ is written as $d_\timestamp =
(\timestamp,v_{A_1},..,v_{A_n})$. Queries over data streams are continuous, i.e. they are
\emph{triggered} when new data arrives. Each
query is composed from one or more \emph{query operators}, which take one or more
streams as inputs and output another stream. We adopt the popular Aurora
query model~\cite{abadi03}, and focus on four operators: Map, Join and Aggregate.
\begin{itemize}
\item \emph{Map}: outputs only the specified attributes. 
\item \emph{Filter}: outputs tuples satisfying a given predicate. 
\item \emph{Join}: takes as inputs two streams $(S_1,S_2)$, two integers $(\ws_1,\ws_2)$ and a join
attribute.  Incoming data are added to the queues of size $\ws_1$ and $\ws_2$, from
which they are joined together.  
\item \emph{Aggregate}: outputs the averages over a sliding window. A sliding window is defined over the timestamp
attribute, with a window size $\ws$ and an advance step $\step$.  
\end{itemize}

\subsubsection{Access Control via Queries.}
\begin{figure}
\centering
{\includegraphics[scale=0.4]{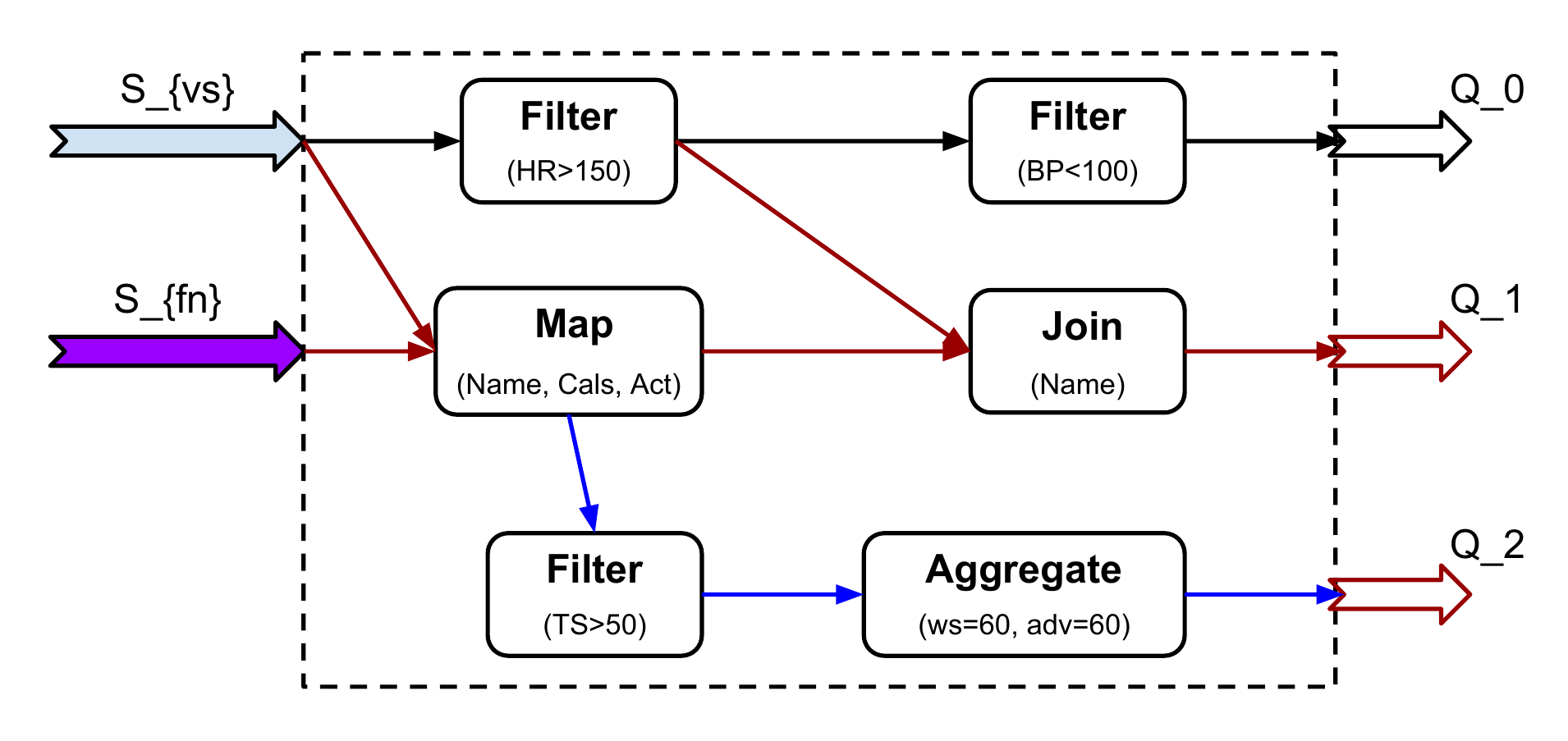}}
\caption{Examples of access control policies via queries}
\label{fig:examples}
\end{figure}
In Streamforce, access control is defined via \emph{views}. As in traditional archival database
systems, views are created by querying the database. In our settings, the access control process
involves two steps. First, the owner specifies a policy by mapping it into a continuous
query. Second, the query is registered to be executed by the cloud, whose outputs are then forwarded
to authorized users. 

The example depicted in Fig.~\ref{fig:examples} includes two streams: $S_\text{\emph{vs}} = $
(\emph{TS, RTime, Name, HR, BP}) and  $S_\textit{fn} = $(\emph{TS, RTime, Name, Cals, Act, Loc}).
$S_\textit{vs}$ contains owner's vital signs as produced by health monitoring devices, where \emph{RTime,
HR, BP} are the real time, heart rate and blood pressure respectively. $S_\textit{fn}$ contains fitness
information, where \emph{Cals, Act Loc} are the number of calories burned, the activity and the
owner's location respectively. Data users could be friends from social network, research institutes
or insurance companies. For a friend, the owner may want to share vitals data when they exceed a certain
threshold ($Q_0$), or average fitness information every hour ($Q_2$). A research institute may be
given a joined view from both streams in order to monitor the individual's vitals during exercises
($Q_1$).  

\subsection{Security Model}
\subsubsection{Adversary Model.}
The cloud is not trusted, in the sense that it tries to learn content of the outsourced data, but it
follows the protocol correctly. This passive (or semi-honest) adversary model reflects the cloud's
incentives to gain benefits from user data  while being bound by the service level agreements. We do
not consider malicious cloud, which may try to break data integrity, launch denial of service
attacks, or compute using stale data.  Security against such attacks is crucial for many
applications, but it is out of the scope of this paper. Data users are considered dishonest, in the
sense that they may proactively try to access unauthorized data. To this end, they may collude with
each other and also with the cloud. 

\subsubsection{Encryptions Model.}
To meet both fine-grained access
control and data confidentiality requirements, we use three different encryption schemes. Proxy
attribute based encryption is used for Map and Filter operators. Second, Join operator is
possible via deterministic encryption. Aggregate is supported by 
sliding-window encryption. This section provides formal definition of these schemes and
their security properties. Detailed constructions and proofs of security are presented in
Section~\ref{sec:encryptionScheme}.

\subsubsection{Deterministic encryption scheme.} $\mathcal{E}_d = (\gen,\enc,\dec)$ is a private-key
encryption scheme, where:
\begin{itemize}
\item $\gen(\kappa)$ generates secret key $\sk$ using security parameter
$\kappa$.
\item $\enc(m,\sk)$ encrypts message $m$ with $\sk$.
\item $\dec(\ct,\sk)$ decrypts the ciphertext. 
\end{itemize}
For any message $m$, $\dec(\enc(m,\gen(\kappa))) = m$. Security of $\mathcal{E}_d$ is defined via the security game consisting of three phases:
\emph{Setup}, \emph{Challenge}, \emph{Guess}
\begin{itemize}
\item Setup: the challenger runs $\gen(.)$.
\item Challenge: the adversary sends to the challenger two messages: $M_0 = (m_{0,0},
m_{0,1},..)$ and $M_1 = (m_{1,0}, m_{1,1},..)$, such that $|M_0| = |M_1|$ and $m_{i,j}$ are all
distinct. The challenger chooses $b \xleftarrow{R} \{0,1\}$, runs $\enc(M_b,\sk)$ and returns
the ciphertext to the adversary. 
\item Guess: the adversary outputs a guess $b' \in \{0,1\}$. 
\end{itemize}
The adversary $\adv$ is said to have an advantage $\adv^\kappa_\mathcal{A} = |Pr[b=b'] -
\frac{1}{2}|$. 
\begin{definition}
$\mathcal{E}_d$ is said to be secure with respect to \emph{deterministic chosen
plaintext attacks}, or {\textbf{Det-CPA secure}}, if the adversary advantage is negligible. 
\end{definition}

\subsubsection{Proxy Attribute-Based Encryption scheme.}
Attribute-Based Encryption (ABE) is a public-key scheme that allows for fine-grained access
control: ciphertexts can only be decrypted if the security credentials satisfy a certain predicate. There are two types
of ABE~\cite{goyal06}: Key-Policy (KP-ABE) and Ciphertext-Policy (CP-ABE). We
opt for the former, in which the predicate is embedded in user keys and the ciphertext contains a
set of encryption attributes.  KP-ABE and CP-ABE can be used interchangeably, but the former is more
data-centric (who gets access to the given data), while the latter is more user-centric (which data
the given user has access to). 

ABE's encryption and decryption are expensive operations. Proxy Attribute Based
Encryption~\cite{green11} (or proxy ABE) is design to aid the decryption process by letting a third party
\emph{transform} the original ABE ciphertexts into a simpler form. It consists of
five algorithms $\mathcal{E}_p =  (\gen,\keygen,\enc,\trans,\dec)$:
\begin{itemize}
\item $\gen(\kappa)$: generates public parameters $\pk$ and master key $\mk$. 
\item $\keygen(\mk,P)$: creates a transformation key $\tk$ and a decryption key $\sk$ for the predicate $P$. 
\item $\enc(m,\pk,A)$: encrypts $m$ with the set of encryption attributes $A$. 
\item $\trans(\tk,\ct)$: partially decrypts the ciphertext using $\tk$. 
\item $\dec(\sk,\ct)$: decrypts the transformed ciphertext using the decryption key.
\end{itemize}
For any message $m$, attribute set $A$, policy $P$, $\mk \leftarrow \gen(\kappa)$, $(\tk,\sk)
\leftarrow \keygen(\mk,P)$, the following holds:
\[
P(A) = 1 \Leftrightarrow \dec(\sk,\trans(\tk,\enc(m,A))) = m
\]
Security of $\mathcal{E}_p$ is defined in~\cite{green11} via a \emph{selective-set security game},
consisting of five phase: \emph{Setup, Query-1, Challenge, Query-2, Guess}:
\begin{itemize}
\item Setup: the challenger executes $\gen(.)$ to generate public parameters. It gives
$\pk$ and an attribute set $A$ to the adversary. 
\item Query-1: the adversary performs either \emph{private key query} or \emph{decryption
query}. In the former, it asks the challenger for the keys of an access structure $T$. The challenger calls
$\keygen$ to generates $(\tk,\sk)$. If $T(A)=0$, it sends both $(\tk,\sk)$ to the adversary. If
$T(A)=1$, it sends $\tk$ to the adversary. For decryption query, the adversary asks the challenger
to decrypt a ciphertext $\ct'$ (which has been transformed using a key $\tk$. The challenger
retrieves the corresponding $\sk$, and calls $\dec(\sk,\ct')$ and sends the result back to the
adversary. 
\item Challenge: the adversary sends two message $m_0,m_1$ of equal length to the
challenger. The challenger chooses $b \xleftarrow{R} \{0,1\}$, computes $\ct \leftarrow \enc(m_b,A)$ and
returns $\ct$ to the adversary. 
\item Query-2: the adversary continues the queries like in Query-1, except that it
cannot ask the challenger to decrypt $\ct$. 
\item Guess: the adversary outputs a guess $b' \in \{0,1\}$. 
\end{itemize}
\begin{definition}
The scheme $\mathcal{E}_p$ is said to be secure with respect to 
\emph{replayable chosen ciphertext attacks}, or \emph{\textbf{R-CCA secure}}, in the selective-set model if the
adversary advantage in the selective-set security game is negligible. 

Modify the security game so that the adversary does not issue decryption queries.
We say that $\mathcal{E}_p$ is secure in the selective-set model with respect to \emph{chosen
plaintext attacks} (or \emph{\textbf{CPA secure}}) if the adversary advantage is negligible. 
\end{definition}

\subsubsection{Sliding-window encryption scheme (SWE).}
$\mathcal{E}_w = (\gen, \enc,\dec)$ is a private-key encryption scheme which allows an
user to decrypt only the aggregate of a window of ciphertexts, and not the individual
ciphertexts.  Let $s(M,\ws)[i]$ and $p(M,\ws)[i]$ be the sum and product of the
$i^{\text{th}}$ window sliding windows (size $\ws$ and advance step $\step = \ws$) over a sequence
$M$. 
\begin{itemize}
\item $\gen(\kappa)$: generates public parameters and the private keys. 
\item $\enc(M=\la m_0,m_1,..,m_{n-1}\ra,W)$: encrypts M using a set of window sizes $W$, whose result is $\ct = \la
c_0,c_1,c_2,..\ra$.  
\item $\dec(\ws, CT,\sk_\ws)$ decrypts $\ct$ for the window size $\ws$ using
the private key $\sk_\ws$. The result is the aggregates of the sliding window, i.e.
$s(M,\ws)[i]$ for all $i$. 
\end{itemize}
Security of $\mathcal{E}_w$ is defined via a \emph{selective-window security game} consisting of
four phases: \emph{Setup}, \emph{Corrupt}, \emph{Challenge}, \emph{Guess}.  
\begin{itemize}
\item Setup: the challenger calls $\gen(.)$ to setup public parameters. It chooses a
value $\ws$ and sends it to the adversary.   
\item Corrupt: the adversary asks the challenger for the private key of a window size
$\ws'$, provided that $\textit{gcd}(\ws,\ws')=\ws$. 
\item Challenge: the adversary picks $M_0  =  \la m_{0,0},
m_{0,1},..,m_{0,n-1}\ra$ and  $M_1 = \la m_{1,0},m_{1,1},..,m_{1,n-1} \ra$, such that
$s(M_0,\ws)[i] = s(M_1,\ws)[i] \label{eq:sw}$ for all $i$ and sends them to the challenger. The adversary also sends a
set of window sizes $W$. The challenger chooses $b \xleftarrow{R} \{0,1\}$, invokes $\enc(M_b,W)$
and forwards the result to the adversary. 
\item Guess: the adversary outputs a guess $b' \in \{0,1\}$. 
\end{itemize}
\begin{definition}
 $\mathcal{E}_w$ is said to be secure with respect to \emph{restricted chosen encrypted window
attacks} (or {\textbf{\emph{Res-CEW secure}}}) in the selective-window model if the adversary's advantage
is negligible.  It is secure with respect to \emph{chosen window attacks} (or {\textbf{\emph{CW
secure}}}) when the Corrupt phase is removed from the game.   
\end{definition}

\subsection{Discussion}
The encryption schemes above have a different definition of security which makes different
assumptions about the adversary's capabilities. R-CCA is the strongest definition, as it assumes
active adversary that has access to the decryption oracles. R-CCA ensures both data integrity and
confidentiality. CPA security assumes a passive (eavesdropping) adversary who only tries to break
the secrecy property of the ciphertext.  CPA security ensures confidentiality, while allowing
meaningful changes to be made on the ciphertext (which is necessary for transformation to work).
Det-CPA is a weaker security level, as it protects data confidentiality only for unique messages.  

Security of the sliding-window scheme $\mathcal{E}_w$ is related to that of secure multi-party
computation, which ensures that no other information is leaked during the computation of a function
except from the final output. Our model is similar, but stronger than the \emph{aggregator oblivious}
model proposed in~\cite{shi11}, since the security game allows for more types of adversarial attacks.
More specifically, \cite{shi11} requires the two message sequences $M_0$ and $M_1$ to have the same
aggregate, but our model requires only the windows (sub-sequences) of $M_0$ and $M_1$ to have the
same aggregate.  Both Res-CEW and CW security  allow for meaningful computations (aggregate) over
ciphertexts. Res-CEW is secure against a weak form of collusion (between users with access to window
sizes which are multiples of each others), whereas CW is not.  

\subsubsection{Access control via Encryption.}
Encryption plays two roles in our system: protecting data confidentiality against untrusted cloud,
and providing access control against unauthorized users. Neither of cloud nor the unauthorized user
have access to decryption keys, hence they cannot learn the plaintexts.
In addition, Res-CEW and CW
security ensure that given access to a window size $\ws$, the user cannot learn 
information of other window sizes (except from what can be derived from its own window). Res-CEW
guarantees access control under weak collusion among dishonest users. 

For access control to be enforced by the cloud, some information must be revealed to the latter.
There exists a trade-off between security and functionality of the query operators that make up
the policies. For Map and Filter policies, the cloud must be able to check if certain attributes are
included in the ciphertexts, which is allowed by CPA security. For Join, the cloud needs to be able 
to compare if two ciphertexts are encryptions of the same message, which requires the encryption to
be deterministic (or Det-CPA secure). For Aggregate, a homomorphic encryption is required, which in
our case means the highest security level is Res-CEW.  

\section{Encryption Scheme Constructions}
\label{sec:encryptionScheme}
\subsection{Deterministic Encryption}
Let $\g$ be a multiplicative group of prime order $p$ and generator $g$. Let $F: \z \times \{0,1\}^*
\to \g$ be a pseudorandom permutation with outputs in $\g$. The scheme $\e_d$ is constructed as follows.
\begin{itemize}
\item $\gen(\kappa)$: $\sk = (k_1,k_2)$ where $k_1,k_2 \xleftarrow{R} \z$.  
\item $\enc(m,\sk)$: $\ct = F(k_1,m)^{k_2}$. 
\item $\dec(\ct,\sk)$:  $m = F^{-1}(k_1,\ct^{\frac{1}{k_2}})$. 
\end{itemize}
\begin{theorem}
Assume that $F$ is a pseudorandom permutation, $\e_d$ is Det-CPA secure
\end{theorem}
\textbf{Proof sketch.} Given any $m_{0,0}, m_{0,1},..$ which are distinct,
 $F(k_1,m_{0,0}), F(k_1,m_{0,1}),..$ are independent and uniformly distributed. As a
consequence, $F(k_1,m_{0,0})^{k_2}, F(k_1,m_{0,1})^{k_2},..$ are also independent and
indistinguishable from random. It follows that $\ct$ is independent from the choice of $M_0$ or
$M_1$, therefore $\text{Pr}[b=b'] = \frac{1}{2}$, or the adversary advantage is $0$. $\qed$

\subsection{Proxy ABE Construction}
Since our adversary model assume passive attackers, we present here the CPA secure construction as
proposed in~\cite{green11} (a R-CCA secure construction can be found in the original paper).
The scheme makes use of \emph{bilinear map} $e:\g_1 \times \g_1 \to \g_2$ where $\g_1,\g_2$ are
multiplicative, cyclic groups of prime order $p$. $e$ is efficient to compute, and $e(u^a,v^b)
= e(u,v)^{ab}$ for $u,v \in \g_1$ and $a,b \in \z$. Its security relies on the
\emph{bilinear decisional Diffie Hellman assumption}: let $g$ be the generator of $\g_1$, for
all $a,b,c,z \xleftarrow{R} \z$, it is difficult to distinguish
$e(g,g)^{abc}$ from $e(g,g)^z$.

\vspace{0.5cm}
$\gen(\kappa)$: generates groups and the bilinear map as the public key. Let $\mathcal{U}$ be
the attribute universe, and $t_1,t_2,..,t_{\mathcal{U}} \xleftarrow{R} \z$. We have: 
\[
\pk = (g,p,\g_1,\g_2,e, T_1 = g^{t_1},..,T_\mathcal{U} = g^{t_\mathcal{U}})
\] 
Let $y \xleftarrow{R} \z$, the master key $\mtext{MK}$ is: $\mtext{MK} = (y,t_1,..,t_{\mathcal{U}})$.  

\vspace{0.5cm}
$\keygen(\mk,P)$: translates $P$ into an access tree, in which the leaf nodes represent
attributes $\mathcal{U}' \subseteq \mathcal{U}$, and the internal nodes represent \emph{threshold gates}. An \textbf{AND} node
corresponds to a $2$-out-of-$N$ gate, an \textbf{OR} nodes to a $1$-out-of-$N$ gate. For each
$k$-out-of-$N$ node $x$ we define a $(k-1)$-degree polynomial $q_x$. Starting from
the root node $r$, defines the polynomial with $q_r(0)=y$. Recursively, for a child node $x$, define $q_x$
such that $q_x(0) = q_{\mtext{parent}(x)}(\mtext{index}(x))$. When $x$ is a leaf node, let $z_u
\xleftarrow{R} \z$ define:
\[
D_x = g^{\frac{q_x(0)}{z_u.t_{a(x)}}}
\] 
where $a(x)$ returns the attribute in $\mathcal{U}$ represented by the leaf node. The
transformation key $\mtext{TK}$ and decryption key $\mtext{SK}$ are defined as follows:
\[
\tk = \{D_x \,|\, a(x) \in \mathcal{U}'\},\  \mtext{SK} = z_u
\]

\vspace{0.5cm}
$\enc(m,\pk,A)$: assume $m \in \g_1$ (or has been mapped from a string to a group
element). Let $s \xleftarrow{R} \z$, the ciphertext is:
\[
\ct = (A, E = m.e(g,g)^{y.s},\  E'=\{T_x^s \,|\, x \in A \})
\]

\vspace{0.5cm}
$\trans(\tk,\ct)$: given the access tree used to generate $\mtext{TK}$, when $x$
is a leaf node, compute:
\begin{align*}
&\texttt{Transform}(x) =\\
&\qquad \left\{
\begin{array}{l l}
e(T_{a(x)},D_{a(x)})=e(g,g)^{\frac{s.q_x(0)}{z_u}} & \mbox{when $a(x) \in A$}\\
\bot & \mbox{otherwise}
\end{array}
\right. 
\end{align*}
When $x$ is a non-leaf node, let $F_z$ be the result from recursive call to
$\texttt{Transform}(z)$ and $z$ is a child node of $x$. Let $S_x$ be the set of $x$'s children
such that $F_z \neq \bot$ for $z \in S_x$. Let $\Delta_{i,S}(x) = \prod_{j \in S, j\neq
i}\frac{x-j}{i-j}$ be the Lagrange coefficient for $i \in \z, S \subseteq \z$. We compute:
\begin{align*}
&\texttt{Transform}(x) = F_x \\
&= \prod_{z \in S_x} F_z^{\Delta_{\mtext{index}(w),S_x'}(0)} \mbox{for } S_x' =
\{\mtext{index}(z) \,|\, z \in S_x\}\\
&= e(g,g)^\frac{s.q_x(0)}{z_u}
\end{align*}
Thus, calling $\texttt{Transform}(r)$ for the root node $r$ results in 
\[\ct' = (U,V) = (E,\texttt{Transform}(r)) = (E,e(g,g)^\frac{y.s}{z_u})\]

\vspace{0.5cm}
$\dec(\sk,\ct')$: the
message can be recovered as: \[
m = \frac{U}{V^{z_u}}
\]

\begin{theorem}[\cite{green11}]
$\mathcal{E}_p$ is \emph{CPA-secure} in the selective-set model.
\end{theorem}

\subsection{Sliding-Window Encryption}
Let $\mathcal{W}$ be the set of all possible window sizes, $\g$ be a multiplicative group of
prime order $p$ and generator $g$. Assuming the message space is a small integer domain, we propose
three different constructions for SWE.

\subsubsection{Construction 1 $(\e^1_w)$:}
masks the plaintext with random values whose sum over the sliding window is the user decryption key. 
\begin{itemize}
\item $\gen(\kappa)$: for all $\ws \in \mathcal{W}$, $\sk_\ws \xleftarrow{R} \z$. 
\item $\enc(M,W)$: for each $\ws \in W$, let $R=(r_0,r_1,..,r_{|M|})$ such that $r_i \xleftarrow{R}
\z$ and $s(R,\ws)[i]=\sk_\ws$. The ciphertext is $\ct = \bigcup_\ws\ct_\ws$ where $\ct_\ws = (g^{m_0+r_0},
g^{m_1+r_1},..)$. 
\item $\dec(\ws,\ct,\sk_\ws)$: extracts $\ct_\ws$ from $\ct$ and compute:\\ $s(M,\text{ws})[i] =
\text{dLog}\left(\frac{p(\ct_\ws,\ws)[i]}{g^{\text{SK}_{\ws}}}\right)$
\end{itemize}
\begin{theorem}
The scheme $\e^1_w$ is Res-CEW secure. 
\end{theorem}
\textbf{Proof sketch.} 
Given input $X=(x_0,x_1,..x_n)$, window size $\ws$ and key $\sk_{\ws}$, define two
distributions $P_0$ and $P_1$ as:
\begin{align*}
&P_0(\ws) = (g^{x_0+r_0}, g^{x_1+r_1},..,g^{x_n+r_n}) \\
&P_1(\ws) = (R_0, R_1,..,R_n)
\end{align*}
where $r_i \xleftarrow{R} \z$ and $R_i \xleftarrow{R} \g$ such that  for all $i$:
\[
\prod_{j \in \ws[i]}R_j = \prod_{j \in \ws[i]}g^{x_j+r_j} = g^{\sk_\ws+s(X,\ws)[i]}
\]
It can be seen that $P_0$ and $P_1$ are indistinguishable (in the information theoretic sense),
because $r_i$ is chosen at random and independently of $x_i$.

Consider the single-window case, i.e. $W = \{\ws\}$. In the security game,  $P_0$ is the
distribution of ciphertext for the input $X$. For input $M_0$, this distribution is
indistinguishable from $D_0 = (R_0,R_1,..R_n)$ where $\prod_{j \in \ws[i]}R_j =
g^{\sk_\ws+s(M_0,\ws)[i]}$. For input $M_1$, the ciphertext distribution is indistinguishable
from $D_1 = (R_0', R_1',..,R_n')$ where $\prod_{j \in \ws[i]}R_j' = g^{\sk_\ws+s(M_1,\ws)[i]}$.
Since\\ $s(M_0,\ws)[i] = s(M_1,\ws)[i]$, $D_0$ and $D_1$ are the same distribution. Therefore,
the adversary can only distinguish the two ciphertext distributions with probability $\frac{1}{2}$.

Consider the case with multiple windows where  $\textit{gcd}(\ws,\ws') = \ws$ for all $\ws'
\in W$. $P_0(\ws)$ and $P_0(\ws')$ are independent, because the random values $r_i, r_i'$ are
chosen independently. They are indistinguishable from $P_1(\ws)$ and $P_1(\ws')$, which are
also independent. Consequently, the combined distribution $(P_0(\ws')|\ws' \in W)$ and
$(P_1(\ws') | \ws' \in W)$ are indistinguishable. Similar to the single-window case above,
using the fact that $s(M_0,\ws')[i] = s(M_1,\ws')[i]$,  the
adversary can only distinguish the two ciphertext distributions with probability $\frac{1}{2}$.$\qed$

\vspace{0.5cm}
\subsubsection{Construction 2 $(\e^2_w)$:} uses an auxiliary encryption scheme to encrypt the
window aggregates directly. 
\begin{itemize}
\item $\gen(\kappa)$: let $\e_\aux = (\gen,\enc,\dec)$ be a CPA-secure asymmetric encryption scheme.
For all $\ws \in \mathcal{W}$, invokes $\e_\aux.\gen(\kappa)$ to generate a key pair $(\pk_\ws,
\sk_\ws)$. 
\item $\enc(M,W)$: $\ct = \bigcup_{\ws \in W} \e_\aux.\enc(\pk_\ws,s(M,\ws)[i])$. 
\item $\dec(\ws,\ct,\sk_\ws)$: extracts $\ct_\ws$ from $\ct$, then computes $s(M,\ws)[i] =
\e_\aux.\dec(\sk_\ws,\ct_\ws[i])$
\end{itemize}
\begin{theorem}
Assuming that $\e_\aux$ is CPA-secure, the construction $\e^2_w$ is Res-CEW secure. 
\end{theorem}
\textbf{Proof sketch.} The proof is similar to that of Theorem~\ref{theorem:window-2}. Because
$\e_\aux$ is CPA-secure, its ciphertext distribution is independent of the input and is
indistinguishable from random. Hence, given $\sk_{\ws'}$ for all $\ws' \in W$ and
$\textit{gcd}(\ws,\ws') = \ws$, both ciphertext distributions $D(M_0)$ and $D(M_1)$ are 
indistinguishable from the following distribution:
\[
(R_0,R_1,..,s(M_0,\ws')[0],s(M_0,\ws')[1],..)
\]
where $R_i \xleftarrow{R} \g$. $\qed$ 

\vspace{0.5cm}
\subsubsection{Construction 3 $(\e^3_w)$:} masks the plaintexts with random values whose sums over
the sliding window are encrypted using another encryption scheme.  
\begin{itemize}
\item $\gen(\kappa)$: the same as in $\e^2_w$.  
\item $\enc(M,W)$: let $R = (r_0,r_1,..,r_{|M|-1})$ where $r_i \xleftarrow{R} \z$, let $\text{CT}_0 =
(g^{m_0+r_0}, g^{m_1+r_1},..)$. For all $\ws \in W$, let $\ct_\ws[i] =
\e_\aux.\enc(\pk_\ws,s(R,\ws)[i])$. Finally, $\ct = \ct_0 \cup \bigcup_{\ws \in W}\ct_\ws$. 
\item $\dec(\ws,\ct,\sk_\ws)$: extracts $\ct_\ws$ from $\ct$, then computes \\
$
s(M,\ws)[i] = \text{dLog}\left(
\frac{p(\ct_0,\ws)[i]}{g^{\e_\aux.\dec(\sk_\ws,\ct_\ws[i])}} \right)
$
\end{itemize}
\begin{theorem}
Assuming that $\e_\aux$ is CPA-secure, the scheme $\e^3_w$ is Res-CEW secure. 
\label{theorem:window-2}
\end{theorem}
\textbf{Proof sketch.} Given $\mathcal{W}$ and $X=(x_0,x_1,..)$, consider the ciphertext distribution:
\[
D_0(X) = (g^{x_0+r_0}, g^{x_1+r_1},..,\ct_{\ws'}[0],\ct_{\ws'}[1],..\,|\, \ws' \in W)
\]
Because $r_i$ is chosen independently from $x_i$, $D_0$ is indistinguishable from $D_1(X) =
(R_0,R_1,..,\ct_{\ws'}[0],\ct_{\ws'}[1]..)$ where $R_i \xleftarrow{R} \g$. 

Let $\mathcal{R} = \{R_0,R_1,..\}$, given $\sk_\ws'$ for all $\ws' \in W$ and $\textit{gcd}(\ws',\ws) = \ws$, $D_1$ becomes:
\begin{align*}
D_2(X) = (&R_0,R_1,..,\\
&\frac{g^{s(X,\ws')[0]}}{p(\mathcal{R},\ws')[0]},\frac{g^{s(X,\ws')[1]}}{p(\mathcal{R},\ws')[1]},..,\\
&\ct_{\ws^*}[0],.. \,|\,\textit{gcd}(\ws^*,\ws)\neq \ws)
\end{align*}
Since $\e_\aux$ is CPA-secure, it follows that $\ct_{\ws*}[i]$ is independent from its input
and indistinguishable from random. That is, $D_2$ is indistinguishable from $D_3$:
\[
D_3(X) =
(R_0,R_1,..,\frac{g^{s(X,\ws')[0]}}{p(\mathcal{R},\ws')[0]},\frac{g^{s(X,\ws')[1]}}{p(\mathcal{R},\ws')[1]},..,T_0,T_1,..)
\]
where $T_i \xleftarrow{R} \g$.

Given the challenge $M_0$ and $M_1$, the ciphertext distribution is $D_3(M_0)$ and $D_3(M_1)$
respectively. Since, $g^{s(M_0,\ws')[i]} = g^{s(M_1,\ws')[i]}$ for all $0 \leq i <
\lfloor\frac{|M_0|}{\ws'}\rfloor$ and $\ws' \in W$ such that $\textit{gcd}(\ws',\ws) = \ws$,
$D_3(M_0)$ is the same as $D_3(M_1)$. Therefore, the adversary can only distinguish the two
distributions with probability $\frac{1}{2}$.  
$\qed$

\section{Secure Query Operators}
\label{sec:secureOperator}
The encryption schemes discussed in previous sections provide the underlying security assurance
for Streamforce. Using encryption directly, access control can be implemented by distributing decryption
keys to the authorized users. Streamforce exposes a higher-level abstraction: system entities deal
only with \emph{secure query operators} which hide the complex and mundane cryptographic details.
This section focuses on the implementation of
the secure operators using the encryption schemes from previous sections. There are three design
components pertaining each operator: (1) how to map the corresponding policy to user decryption key, (2)
how to encrypt the data at the owner, (3) how the transformation at the cloud is done. Many
fine-grained policies can be constructed by using one of these operators directly. We also describe the
design for combining these operators to support more complex policies.  
 
\subsection{Map}
This operator returns data tuples containing only attributes in a given set $\bb$. We use $\e_p$ to
implement this operator. First, $\e_p.\gen(.)$ is invoked to setup the public parameters and master
key $\mk$. The user decryption key is created by 
{\small $\e_p.\keygen(\mk,\texttt{P-Map}(\bb))$}, where: 
{\small
\[\texttt{P-Map}(\bb) = (`\att=B_1\textrm' \cap `\att=B_2\textrm'
\cap..)
\]
} The owner encrypts using:
{\small
\[
\texttt{Enc-Map}(d_\timestamp)
 = \big(\timestamp,\e_p.\enc(v_{A_1},\{`\att=A_1\textrm'\}),
\e_p.\enc(v_{A_2},\{`\att=A_2\textrm'\}), ..\big)
\]
} When the ciphertext $\ct$ arrives at the cloud, it is transformed using $\e_p.\trans(\tk,\ct)$
before being forwarded to the user. 

This implementation achieves the same level of security as $\e_p$, i.e. CPA security in the selective set
model. The storage cost is $O(|\ab|.\text{log}(p))$ bits per data tuple, where
$p$ is the size of $\g$. 

\subsection{Filter}
Let  $\fa$ be the set of \emph{filter attributes}. A filter \emph{predicate} is
defined by a tuple $(A,k,\op)$ in which $k \in \mathbb{Z}, A \in \ab, \op \in
\{`=\textrm', `\leq\textrm', `\geq\textrm', `\text{mod }p\textrm'\}$. The predicate returns  
true when $(A\  \op \ k)$ returns true. Let $\lambda(A,v,b)$ be the bag-of-bit representation of
$v_A$ in base $b$, as explained in~\cite{goyal06}. In particular, assuming $v = (y_0y_1..y_{m-1})_b$, we have:
\[
\lambda(A,v,b) = 
\begin{Bmatrix}
`\att_{A,b,0} = (xx..xy_0)_b\textrm'\\
`\att_{A,b,1} = (xx..y_1x)_b\textrm'\\
\cdots \\
`\att_{A,b,m-1} = (y_{m-1}x..x)_b\textrm'
\end{Bmatrix}
\]
Let $\mathbb{P}$ be a set
of primes, define
{\small
\[
\textit{AS}(A,v) = \bigcup_{b \in \mathbb{P}}\lambda(A,v,b) \cup \{`\att_A = \text{
don't care }\textrm'\}
\]} as the set of encryption attributes representing the value $v$. 

\begin{algorithm}
\scriptsize
\condge$(A,s,i)$:\\
$\ $ if  $(i==1) \text{ AND } s[0]=='0')$\\
$\qquad$ \textbf{return} $`\att_{A,2,1}=xx..1x\textrm'$\\
$\ $ else\\
$\quad$ if $(s[i]==`0\textrm')$\\
$\qquad$ \textbf{return} $`\att_{A,2,i}=x..\underbrace{1x..x}_{i}\textrm' \ \cup $ \condge$(s,i-1)$ \\
$\quad$ else if $(i==0)$ \textbf{return} $`\att_{A,2,0}=x..x1\textrm'$\\
\vspace{0.2cm}
$\quad$ else \textbf{return} $`\att_{A,2,i}=x..\underbrace{1x..x}_{i}\textrm'\ \cap $ \condge$(s,i-1)$ \\
\caption{Generate policy condition when op is $\ge$.}
\label{alg:ge}
\end{algorithm}
\begin{algorithm}
\scriptsize
\condle$(A,s,i)$:\\
$\ $ if $(i==0)$ \textbf{return} $`\att_{A,2,0}=x..x0\textrm'$\\
$\ $ else\\
$\quad$ if $(s[i]==`0\textrm')$\\
$\qquad$ \textbf{return} $`\att_{A,2,i}=x..\underbrace{0x..x}_{i}\textrm'\ \cap $ \condle$(s,i-1)$ \\
$\quad $ else \textbf{return} $`\att_{A,2,i}=x..\underbrace{0x..x}_{i}\textrm'\ \cup $
\condle$(s,i-1)$ \\
\caption{Generate policy condition when op is $\le$}
\label{alg:le}
\end{algorithm}

Denote $D(A,k,\op)$ as the policy corresponding to the predicate $(A,k,\op)$. 
When $\op$ is $`=\textrm'$, $D(A,k,`=\textrm') \leftarrow \bigcap_{P \in \lambda(A,k,2)} P$. When
$\op \in \{\leq,\geq\}$, $D(A,k,\op) \leftarrow \texttt{cond\_le}(A,(y_{m-1}..y_0)_2, m-1)$ or
$D(A,k,\op) \leftarrow \texttt{cond\_ge}(A,(y_{m-1}..y_0),m-1)$ where \texttt{cond\_le} and
\texttt{cond\_ge} are detailed in Alg.~\ref{alg:le} and Alg.~\ref{alg:ge} respectively. 

When $\op$ is $`\text{mod }p\textrm'$, we consider three cases. 
\begin{itemize}
\item If $p \in \mathbb{P}$: $D(A,k,`\text{mod }p\textrm') \leftarrow `\att_{A,p,0} = (xx..xk')_p\textrm'$ where $k' = k \text{ mod
} p$. 
\item If there exists $q \in \mathbb{P}$ and $p = q^t$ for some $t$. Let $k' =
(y_{t-1}..y_1y_0)_q$ be the representation of $k' = k \text{ mod  }p$ in base $q$. Let
\[
D(A,k,`\text{mod }p\textrm') = \bigcap 
\begin{pmatrix}
`\att_{A,q,0} = (xx.xy_0)_q\textrm' \\
\cdots \\
`\att_{A,q,t-1}=(x..\underbrace{y_{t-1}x..x}_{t})_q\textrm'\\
\end{pmatrix}
\]
\item $p=q_1^{t_1}q_2^{t_2}..q_m^{t_m}$ for $q_i \in \mathbb{P}$ and some values of $t_i$. Then, we have
$D(A,k,`\text{mod }p\textrm') \leftarrow \bigcap_{i}D(A, k \text{ mod } q_i^{t_i}, \text{ mod } q_i^{t_i})$. 
\end{itemize}
The user decryption key is generated by
{\small $\e_p.\keygen\Big(\mk,\texttt{P-Filter}\big(\{(A,k,\op)\}\big)\Big)$}, where
{\small
\[
\texttt{P-Filter}\big(\{(A,k,\op)\}\big) = \bigcap_{A \in \fa} D(A,k,\op)
\]
}
The owner encrypts data using: 
{\small
\[
\texttt{Enc-Filter}(d_\timestamp) = \left(\{v_A\,|\, A \in \fa \},\, \e_p.\enc\Big(d, \bigcup_{A \in
\fa}\text{AS}(A,v_A)\Big)\right)
\]
} When the ciphertext $\ct$ arrives at the cloud, the latter transforms it using
$\e_p.\trans(\tk,\ct)$ and forwards the result to the user.
 
Similar to $\map$, this operator uses proxy ABE scheme directly, therefore it has CPA
security in the selective set model. The storage cost per ciphertext is
$O(|\text{FA}|.|\mathbb{P}|.\text{log}(p))$ bits, which grows with the size of
$\mathbb{P}$. The bigger the size of $\mathbb{P}$, the more policies of the type $`\text{ mod p
}\textrm'$ can be supported, but at the expense of more storage overhead. Notice that values of 
filtering attributes are exposed to the cloud in the form of encryption attributes, thus the data owner should
only use non-sensitive attributes, such as $\TS$, for the set $\fa$.

\subsection{Join}
Let $J$ be the join attributes of two streams $S_1,S_2$. We assume that the join
operator returns all data attributes (more complex cases are discussed in
Section~\ref{subsec:combination}).  We use a combination of proxy ABE scheme $\e_p$ and deterministic
scheme $\e_d$. Initially, the two owners of $S_1,S_2$ invoke $\e_d.\gen(.)$ in a way that
satisfies two conditions: (1) both end up with the same group $\g$ and pseudorandom function
$F$; (2) $\sk_1 = (k_{1,1},k_{1,2})$ and $\sk_2 = (k_{2,1},k_{2,2})$ are the two secret keys such
that $k_{1,1} = k_{2,1}$.  

The user decryption for stream $i$ is {\small $\big(k_{i,2},\,
\e_p.\keygen(\mk,\texttt{P-Join}(J))\big)$}, where
{\small
\[
\texttt{P-Join}(J) = `\att = J\textrm'
\]
}
The owner encrypts using:
{\small
\[
\texttt{Enc-Join}(d_\timestamp,J) = (U,V) =  \Big(\e_p.\enc(d,`\att = J\textrm'), \e_d.\enc(v_{J})\Big)
\]}
The user who received both $k_{1,2}$ and $k_{2,2}$ computes $(z_1 = \frac{s}{k_{1,2}}, z_2 =
\frac{s}{k_{2,2}})$ where $s \xleftarrow{R} \z$ and sends it to the cloud. When two ciphertexts 
$(U_1,V_1)$ and $(U_2,V_2)$ arrive at the cloud, it checks if $V_1^{z_1} = V_2^{z_2}$.  If true,
the ciphertexts can be joined. The cloud then performs $\e_p.\trans(\tk_1,U_1)$,
$\e_p.\trans(\tk_2,U_2)$ and forwards the results to the user. 

Because $\e_d$ is Det-CPA secure, the cloud can learn if the encryption of $v_J$ is the same as
in both streams. But this check is only possible if the user requests it (by
sending $z_1$ and $z_2$ to the cloud). Other attributes in $d_\timestamp$ are protected with CPA
security by $\e_p$. The storage requirement is $O(\text{log}(p))$
bits per data tuple, because $\e_d.\enc(.)$ produces a group element and $\e_p$ encrypts the entire
data tuple with only
one encryption attribute. 

\subsection{Aggregate (Sliding Window)}
\begin{figure}
\centering
\subfloat[]{\includegraphics[scale=0.62]{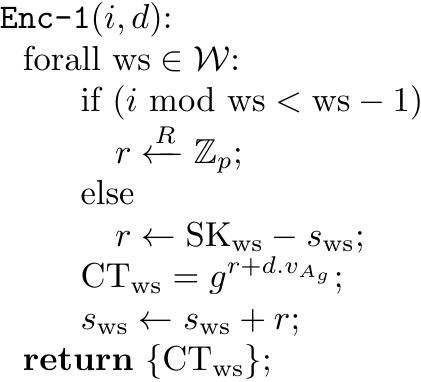}}
\hspace{0.15cm}
\subfloat[]{\includegraphics[scale=0.62]{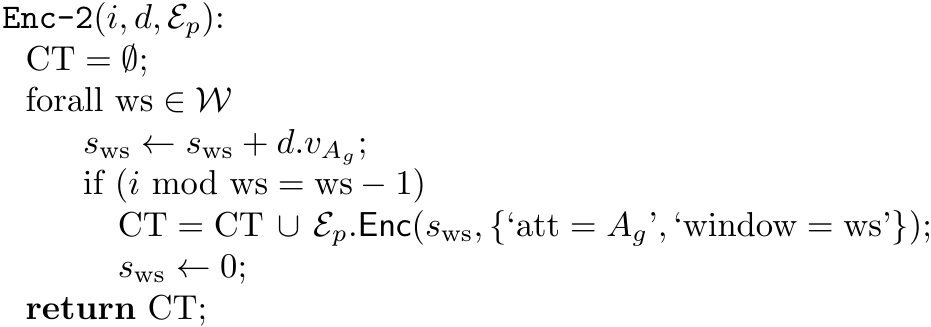}}
\hspace{0.25cm}
\subfloat[]{\includegraphics[scale=0.62]{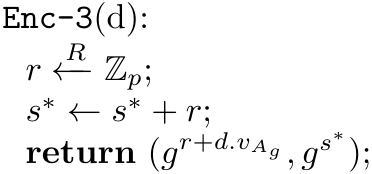}}
\caption{Encryption used for aggregate protocols}
\label{fig:algs}
\end{figure}
In Streamforce, sliding windows are based on timestamp attribute $\TS$, with
advance steps being the same as the window sizes. Let $A_g$ be the aggregate attribute, over which
the sums are computed. In the
following, we present three implementations for this operator, and discuss their trade-offs at the
end. 
 
\paragraph{\textbf{Agg-1}.}
The owner first encrypts data using $\e^1_w$, the ciphertext is then encrypted with $\e_p$. The
user decryption key is {\small $\big(\sk_\ws,\, \e_p.\keygen(\mk, \texttt{P-Agg1}(\ws,A_g))\big)$},
where $\sk$ is the secret key generated by $\e^1_w.\gen(.)$, and 
{\small
\[ \texttt{P-Agg1}(\ws,A_g) = `\att = A_g\textrm' \,\cap\,
`\textit{window}=\ws\textrm'
\]
}
To encrypt $d_\timestamp$, the owner first
executes $\texttt{Enc-1}(\timestamp,d_\timestamp) \to \{\ct_\ws\}$ as shown in
Fig.~\ref{fig:algs}[a], then computes: {
\small
\[
\texttt{Enc-Agg1}(d_\timestamp,\{\ct_\ws\}) = \Big(\e_p.\enc(d, \{`\textit{window}=1\textrm'\})
 ,\ \bigcup_{\ws \in \mathcal{W}}\e_p.\enc(\ct_\ws, \{`\att = A_g\textrm',
`\textit{window}=\ws\textrm'\})\Big)
\]
} 
For every window size $\ws$, the cloud maintains a buffer of size $\ws$.  The incoming ciphertext
$\ct$ is transformed using $\e_p.\trans(.)$, and the result is added to the buffer. Once the buffer
is filled, the cloud computes the product of its elements, sends the result to the user and clears
 the buffer. 

\paragraph{\textbf{Agg-2}.}
This implementation uses $\e^2_w$ with $\e_p$ as the auxiliary encryption scheme. The
owner itself computes the window aggregates and encrypts the result using $\e_p$. User decryption key is
{\small $\e_p.\keygen(\mk,\texttt{P-Agg2}(\ws,A_g))$}, where: 
{\small
\[\texttt{P-Agg2}(\ws,A_g) = `\att=A_g\textrm' \,\cap\,
`\textit{window}=\ws\textrm'
\]
}
To encrypt $d_\timestamp$, the owner first executes $\texttt{Enc-2}(\timestamp,d_\timestamp,\e_p)
\to \ct$ as shown in
Fig.~\ref{fig:algs}[b], then the ciphertext is computed as: 
{\small
\[
\texttt{Enc-Agg2}(d_\timestamp) = \Big(\e_p.\enc(d_\timestamp,\{`\textit{window}=1\textrm'\}), \, \ct\Big)
\]
}
At the cloud, the ciphertexts for a window aggregate are of the same form as for a normal data tuple.
The cloud simply invokes $\e_p.\trans(\ct_\ws, \tk_\ws)$ and forwards the results to the user. 

\paragraph{\textbf{Agg-3}.}
This implementation uses $\e^3_w$  with $\e_p$ as the auxiliary encryption scheme. The user key is
{\small $\e_p.\keygen(\mk,\texttt{P-Agg3}(\ws,A_g))$}, where:
{\small
\[ \texttt{P-Agg3}(\ws,A_g) = `\att = A_g\textrm' \ \cup\ D(\TS,\ws-1,`\text{mod } \ws\textrm')
\]
}
To encrypt $d_\timestamp$, the owner first computes $\texttt{Enc-3}(d_\timestamp) \to (U,V)$ as
shown in Fig.~\ref{fig:algs}[c] where $s^* \xleftarrow{R} \mathbb{Z}_p$ is a public parameter. The
ciphertext is: 
{
\small
\[\texttt{Enc-Agg3}(d_\timestamp)  = \Big(\timestamp,\, \e_p.\enc(v_{A_g},\{`\textit{window}=1\textrm'\}),
\ \e_p.\enc(U,\{`\att = A_g\textrm'\}),\, \e_p.\enc\big(V, \text{AS}(\TS,i)\big)\Big)
\]
}
The cloud maintains a $\ws$-size buffer, and a variable $X$ whose initial
value is $(g^{s^*},1)$. For the incoming ciphertext $\ct = (\timestamp,U,V,Z)$, the cloud performs
$\e_p.\trans(\tk,V)$ and adds the result to the buffer. Once the buffer is filled (at index $\timestamp$), the
cloud computes the product $U'$ of the buffer elements and clears the buffer. Next, it computes $V' \leftarrow
\frac{\e_p.\trans(\tk,Z)}{X}$,  and then assign $X \leftarrow V'$. Finally, it sends $(U',V')$
to the user, at which the sum is decrypted as:
$\text{dLog}\left(\frac{\e_p.\dec(\sk,U')}{{\e_p.\dec(\sk,V')}}\right)$  

\paragraph{\textbf{Discussion.}}
Unlike $\map$, $\filter$ and $\join$, the \texttt{Aggregate} operator requires more effort from the cloud, i.e.
multiplication of ciphertexts. Since it uses $\e_p$ as the final layer of encryption, it achieves
CPA security with respect to the cloud. In all three implementations, the transformed ciphertexts
received by the user are data encrypted with $\e^1_w$, $\e^2_w$ or $\e^3_w$. As discussed in the
previous section, these schemes achieve Res-CEW security, therefore the user learns nothing more
than the aggregate values.  

While both security with respect to the cloud and access control property are the same for
the three protocols, their main differences lie in their flexibility and storage cost:
\begin{itemize}
\item \emph{Flexibility}: Agg-1 and Agg-2 support a fixed set of window sizes, as
defined by $W$. Agg-3, however, specifies a set of prime $\mathbb{P}$ and is able to
support any window size $\ws$ which can be factorized into the form $(\ws =
q_1^{t_1}.q_2^{t_2}..)$ for all $q_i \in \mathbb{P}$ and $t_i \in \mathbb{N}$. For instance,
with $\mathbb{P}=\{2,3,5\}$, Agg-3 can support any window size in
$\{2,3,5,4,6,8,9,20,23,..\}$. Finally, Agg-3 allows the data owners to specify windows
starting from arbitrary positions, as opposed to the fixed starting position $\{0, \ws,
2.ws,..\}$
for windows of size $\ws$. For this usage, however, the security model must be extended to deal
with collusion not only from users whose policies permit access for different window sizes, but
also from users having access to the same window sizes but from different starting positions.  

\item \emph{Storage cost}: In Agg-1, each data tuple needs to be encrypted $W$ times, each
for a different window size. Hence, its cost is $O(|W|.\text{log}(p))$ bit per encrypted data
tuple. In Agg-2, encrypting data tuple $d_i$ may follow by encryptions of window sums for
windows that ends at $i$. In the worst case, the cost for encryption of $d_i$ is
$O(|W|.\text{log}(p)$. Notice that even though both protocols have the same asymptotic cost,
bound for memory cost, the cost incurred by Agg-2 is much cheaper in practice, because
most of the time $d_i$ requires only one encryption, whereas Agg-1 always requires $W$
encryptions for all $d_i$. In Agg-3, even though each $d_i$ requires only two
encryptions, each encryption requires bigger storage for all the attributes in
$\textit{AS}(\TS,i)$. Hence, its memory cost is $O(|\mathbb{P}|.\text{log}(p))$.   
\end{itemize}

It can be seen that there is a trade-off between flexibility and storage
overhead. In particular, when the owner wishes to support a small number of
windows, Agg-2 is a better choice among the three. However, when more flexible
windows are required, Agg-3 may have a better trade-off between flexibility and storage
cost. Our experimentation with Streamforce in Section~\ref{sec:evaluation} suggests that this is
indeed the case. 

\subsection{Combining Multiple Operators}
\label{subsec:combination}
Each operator presented above can be used by itself to support a wide range of fine-grained
policies. However, many more policies can be supported when two or more of these operators are
combined together. In the following, we show how to implement such high-level combinations.  

\paragraph{\textbf{Map and Filter.}}
The user decryption key is generated by combining the Map and Filter key, i.e.
$\e_p.\gen\big(\mk,\texttt{P-Map}(\bb),\texttt{P-Filter}(\{A,k,\op\})\big)$. The owner encrypts
using:
{\small
\[
\texttt{MF-Enc}(d_\timestamp) = \big(\timestamp,\, \{v_A \,|\, A \in \fa\},\, \e_p.\enc(v_{A_1}, A^*_1),\, 
\e_p.\enc(v_{A_2}, A^*_2), .. \big)
\]
} where $A^*_i = \{`\att=A_1\textrm'\} \, \cup \, \bigcup_{A \in \fa} AS(A,v_a)$
This operator is CPA secure, and the storage cost is
$O(\ab.|\fa|.|\mathbb{P}|.\text{log}(p))$ bits per data tuple. 

\paragraph{\textbf{Map, Filter and Join.}}
This operator allows the cloud to join two encrypted streams only when filter conditions on each
stream are met. The user decryption key is made up of the Map-Filter key and the Join key, i.e:
{\small
\[
\Big(k_{i,2},\, \e_p.\keygen\big(\mk,\texttt{P-Join}(J)\big),\,
\e_p.\gen\big(\mk,\texttt{P-Map}(\bb),\texttt{P-Filter}(\{A,k,\op\})\big)\Big)
\]
}
The data owner encrypts using:
{\small
\[
\texttt{MFJ-Enc}(d_\timestamp) = \Big(\{v_A\,|\, A \in \fa\},\, \texttt{MF-Enc}(d_\timestamp),\, \e_p.\enc\big(\e_d.\enc(v_J)\Big)
\]
}
This operator is Det-CPA secure, and its storage cost is dominated by the cost of $\texttt{MF-Enc}$,
which is $O(\ab.|\fa|.|\mathbb{P}|.\text{log}(p))$ bits per data tuple. 

\paragraph{\textbf{Filter and Aggregate.}}
We assume that each
sliding window contains only \emph{continuous} elements, i.e. $\{d_\timestamp, d_{\timestamp+1},..,
d_{\timestamp+\ws-1}\}$.  Therefore, combining Filter and Aggregate only applies to filtering
conditions of the form $(\TS,k,`\ge\textrm')$ for $k \in \mathbb{Z}$. When Agg-2 is used, the user decryption key is:
{\small
\[
\Big( \e_p.\keygen\big(\mk,\texttt{P-Agg2}(\ws,A_g)\big),\,
\e_p.\keygen\big(\mk,\texttt{P-Filter}(\{\TS,k,`\ge\textrm'\})\big)
\Big)
\]
}
To encrypt $d_\timestamp$, the owner simply invokes $\texttt{Enc-Agg2}(d_\timestamp)$. Thus, this
operator has the same security level as that of Agg-2: CPA security against the cloud and
Res-CEW security against users. The memory cost is $O(|W|.\text{log}(p))$ bits per
data tuple.  

When Agg-3 is used, the user key is created by $\e_p.\keygen(\mk,\texttt{FA-Policy}(\ws,A_g,x))$
where:
\begin{align*}
\texttt{FA-Policy}(\ws,A_g,x) = & \big(`\att=A_g\textrm' \,\cap\,D(\TS,x,`\ge\textrm')\big)\\
& \cup \, \big(`\att=\TS\textrm' \,\cap\, D(\TS,x-1,`\ge\textrm') \, \cap\, D(\TS,x-1,`\text{mod }
\ws\textrm')\big)
\end{align*} 
The encryption of $d_\timestamp$ is:
\begin{align*}
\texttt{FA-Enc}(i,d_i) = & \Big(i,\, \e_p.\enc(v_{A_g},\{`\textit{window}=1\textrm'\}),\\
& \quad \e_p.\enc(U,\{`\att = A_g\textrm'\}\,\cup\,\textit{AS}(\TS,i)),\\
& \quad \e_p.\enc(V, \{`\att = \TS\textrm'\}\,\cup\,\textit{AS}(\TS,i))\Big)
\end{align*}
where $(U,V)$ is output of $\texttt{Enc-3}(d_\timestamp)$. 
The implementation of Aggregate box is modified slightly as follows. The cloud will only start
transforming and filling the buffer when $i=x$. In addition, when $x=0$, the initial value of $X$
is $(g^{s^*},1)$ as before. When $x>0$, the cloud waits to receive the ciphertext
$(x-1,U,V,Z)$ and then initializes $X \leftarrow \e_p.\trans(\tk,Z)$. The memory cost is
$O(|W|.\text{log}(p))$.




\section{Prototype and Evaluation}
\label{sec:evaluation}
\subsection{Implementation and Benchmark}
\begin{table}
\centering
\hspace*{-1cm}
\scriptsize
\begin{tabular}{|c|l|l|}
\hline
\textbf{Policy} & \textbf{Description} & \textbf{Esper query} \\ \hline
T1 & select certain stock & select * from StockEvent(stockId=x)\\ \hline
T2 & stock within timestamp range & select * from StockEvent(stockId=x, $ y < \timestamp < z$)\\ \hline
T3 & stock within time interval & select * from StockEvent(stockId=x, $y < \text{hour} < z$)\\ \hline
T4 & stock every fixed interval & select * from StockEvent(stockId=x, $\timestamp \% x = y$)\\ \hline
\multirow{2}{*}{T5} & aggregate (Agg-1,3) & select price('ws=l'), volume('ws=l') from
StockEvent(stockId=x).win:length\_batch(y) \\ 
 & aggregate (Agg-2) & select price('ws=l'), volume('ws=l') from StockEvent(stockId=x)\\
\hline
\multirow{3}{*}{T6} & \multirow{3}{*}{join price} & select * from StockEvent(stockId=x, $y<
\timestamp < z$)
$\quad$\emph{//output StockJoinEvent stream} \\ 
& & select * from StockJoinEvent(policyId=p).win:length($l_1$) as $s_1$,\\ 
& & $\qquad$ StockJoinEvent(policyId=p).win:length($l_2$) as $s_2$ where
$s_1$.price('det')=$s_2$.price('det') \\ \hline
\end{tabular}
\caption{Access control policies}
\label{tab:policies}
\vspace{-0.5cm}
\end{table}

We implement a prototype of Streamforce~\cite{streamforce} over
Esper\footnote{\url{esper.codehause.org}}. Esper is an open source stream processing engine which
can process millions of data items per second. One can register a continuous query to Esper, then
implement a \emph{listener} that processes the output stream. In Streamforce, policies are translated
into queries (Table~\ref{tab:policies}), and transformations for each policy are done at the corresponding listener.
We leverage Esper to manage policies, to quickly process the ciphertext streams (i.e. direct the ciphertext to
the correct listener), and to handle the complex join operation. We use OpenSSL's AES
implementation for deterministic encryption scheme, while proxy ABE and sliding window schemes are
implemented by extending the KP-ABE library~\cite{libcelia}.  

We create a benchmark containing stock market data of the scheme:

{$\qquad \qquad$ \emph{StockEvent = (TS, hour,  stockId, price,volume)}}\\
in which \emph{hour} values are in $[0,24)$ while
\emph{price, volume} values are in $[0,100)$.  Each stream is identified by its \emph{stockId}. The
benchmark data contains 1 million encrypted data tuples belonging to 100 streams, which is over 100GB
in size and is available on request. We generate different types of policies, as listed in Table~\ref{tab:policies} 
which  also shows how the policies are translated into Esper
queries. Notice that when Agg-1 or Agg-3 implementation is used, the query involves
Esper's window operator because we rely on Esper to maintain the window's buffer. In contrast,
Agg-1 requires no window since the cloud only transforms individual ciphertexts. Join policies
use Filter-Join operators (the Filter conditions are similar to those of T2,T3 and T4 policies), and
involves two steps: the first transforms the input stream into \emph{StockJoinEvent} stream
containing the deterministic encryption of the join attribute, the second takes two
StockJoinEvent streams and produces join outputs.   

We first benchmark individual cost of various operations at the owner, the cloud and the user by
measuring their execution time. Next, we evaluate system performance in terms of
\emph{\textbf{throughput}} and \emph{\textbf{latency}}. Throughput is quantified by
the number of unique data tuples processed by the system per second.  For join policies, however, it is
measured as the number of join outputs processed per second. Latency is determined from the time a data
tuple enters Streamforce to the time it is sent to the user. This metric includes both queuing time and
transformation time. Our experiments were carried out on Amazon's EC2 instances, with $8$ window sizes
($\{2,4,8,..,256\}$) and maximum of $100$ policies (mixture of all different types) per stream.  
  
\subsection{Experiment Results}
\begin{table}
\footnotesize
\centering
\begin{tabular}{|c|c|c|c|}
\hline
\textbf{Type} & \textbf{Throughput} & \textbf{Latency at saturation (ms)}  & \textbf{Latency at rate 1 tuple/sec}\\ \hline
{m1.large} & $125.58 \ (\pm 3.32)$ & $781$  & $9.12 \ (\pm 0.05) $\\ \hline
{m2.xlarge} & $160.26 \ (\pm 5.88)$ & $628$ & $6.95 \ (\pm 0.04) $  \\ \hline
{m3.xlarge} & $248.28 \ (\pm 8.54)$ & $567$ & $5.70 \ (\pm 0.11) $ \\ \hline
\end{tabular}
\caption{Performances for different cloud instances}
\label{tab:ec2types}
\end{table}
We start with a simple workload consisting of one stream and one T1 policy.  We run the workload on
different types of EC2 instances with different capacity, including (from small to large):
\emph{m1.large, m2.xlarge} and \emph{m3.xlarge}. We vary the data rate, and observe the system
performance at saturation point. As seen in Tab.~\ref{tab:ec2types}, \emph{m3.xlarge} achieves the
best performance, with throughput of $249$ (tuples/sec) and latency of $567ms$ (at $99^{\text{th}}$
percentile). In contrast, \emph{m1.large} and \emph{m2.xlarge} have lower throughputs at $125$ and
$160$ (tuples/sec), and higher latency at $781ms$ and $628ms$.  This is because \emph{m3.xlarge}
have more CPU power than other types ($3.5$ CPU units as compared to $2.5$ and $2$ units). The
remaining results presented below are from experiments running on \emph{m3.xlarge} instances. 

\begin{figure}
\centering
{\includegraphics[scale=0.4]{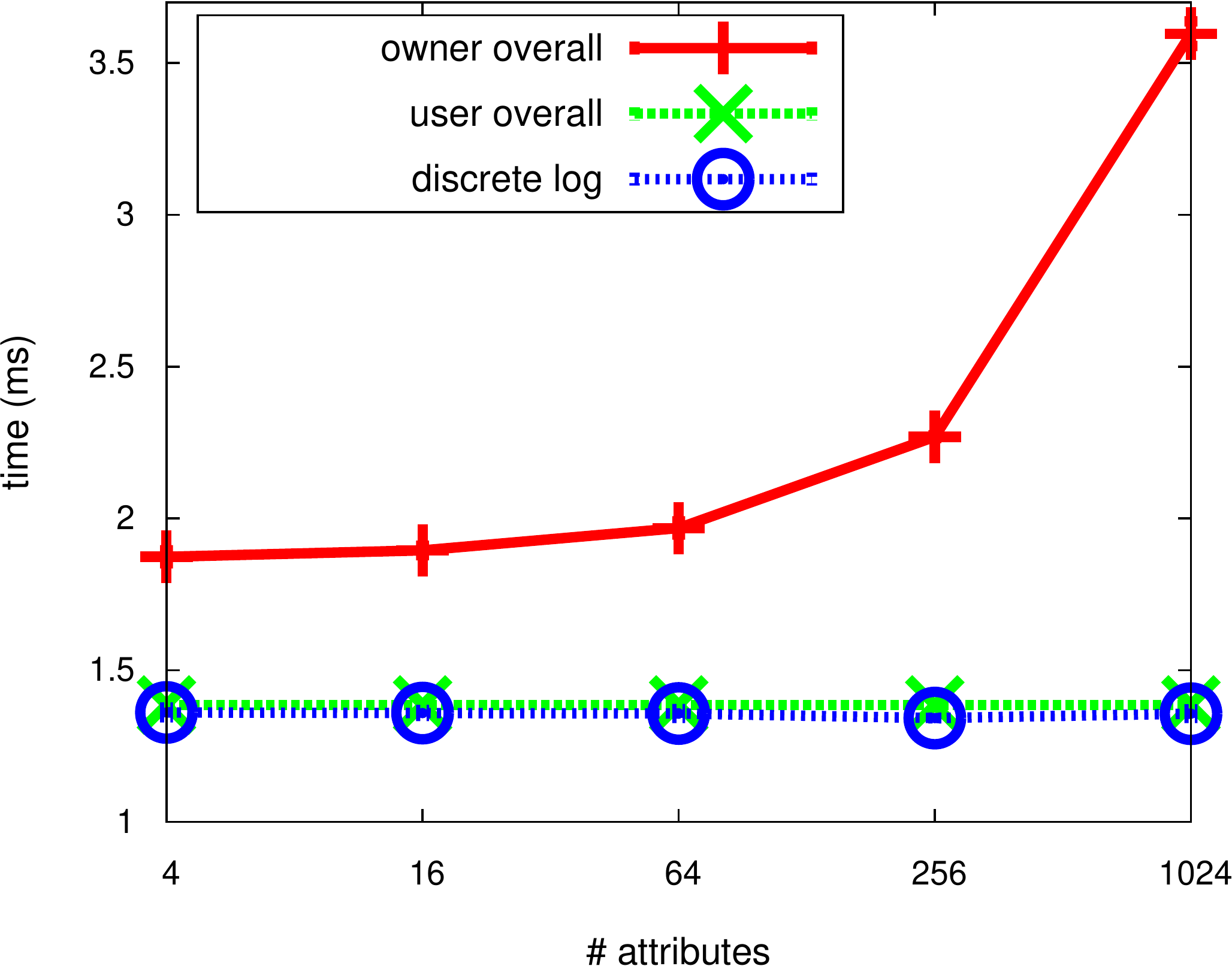}}
\caption{Initialization time}
\label{fig:init}
\end{figure}
When the system first starts, the owners and users have to initialize the cryptographic sub-systems
(running $\gen(.)$, among other things). This one-off cost consists of a constant cost for
pre-computing discrete logarithms, and a variable cost depending on the number of encryption
attributes. Fig.~\ref{fig:init} shows that even with $1024$ encryption attributes, this
initialization process takes less than $3.5s$. 

\begin{figure}
\hspace*{-1.5cm}
\centering
\subfloat[Encryption (at owner)]{\includegraphics[scale=0.35]{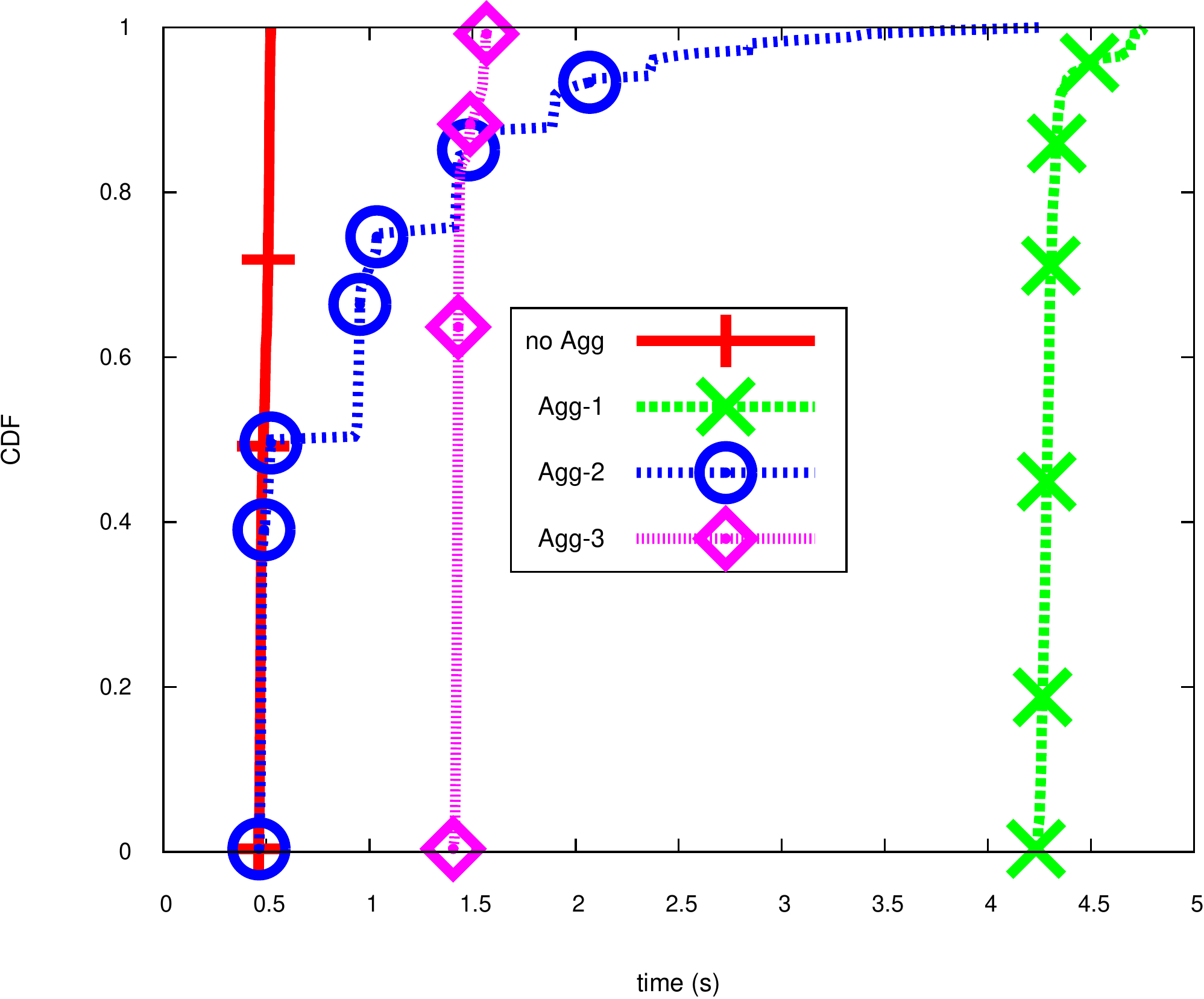}}
\subfloat[Transformation]{\includegraphics[scale=0.35]{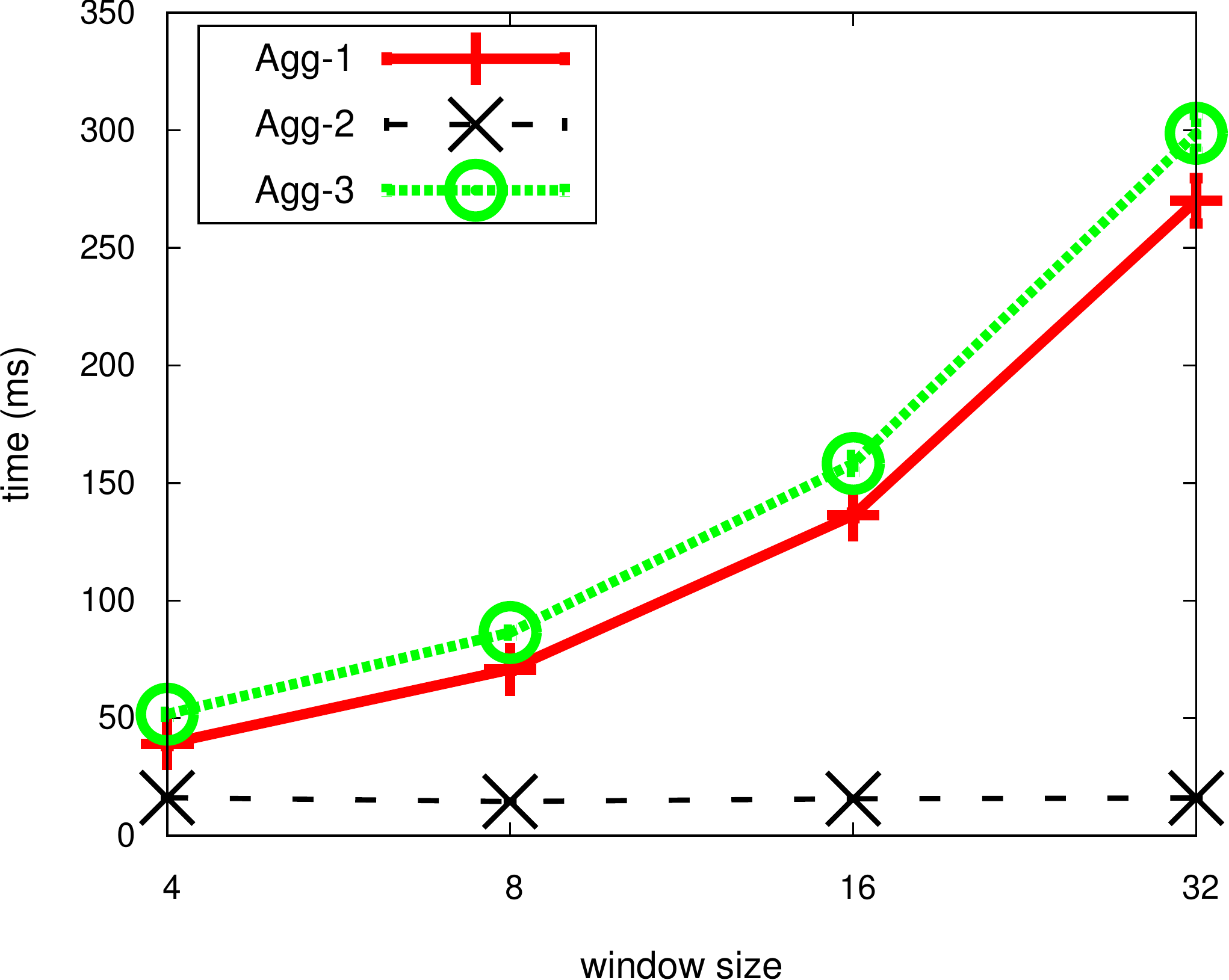}}
\caption{Transformation for aggregate policies}
\label{fig:agg}
\end{figure}
Fig.~\ref{fig:agg}[a] shows the cost of encryption per data tuple at the owner. If the owner does
not allow for aggregate policies, it is relatively constant at approximately $0.5s$. The cost for
supporting Agg-1 is the largest (over $4s$), since the owner has to encrypt the data multiple
times (one for each window size). Agg-3 is also more expensive, since two extra columns are
encrypted for each tuple. The cost of Agg-2 stays low for most of the time (its maximum value
is still as high as of that of Agg-1). This agrees with our analysis in
Section~\ref{sec:secureOperator}, i.e.  most of the time the owner incurs no extra encryption per
tuple, but in the worst case it has to do 8 encryptions per tuple. Fig.~\ref{fig:agg}[b] compares
the transformation costs at the cloud for different implementations of the aggregate operator. It can
be seen that for Agg-2 the cost is constant, whereas for others it is linear with
the size of the window. This is because for Agg-1 and Agg-3, the cloud needs to transform many
ciphertexts and multiply them to get the average.  

\begin{figure}
\centering
{\includegraphics[scale=0.4]{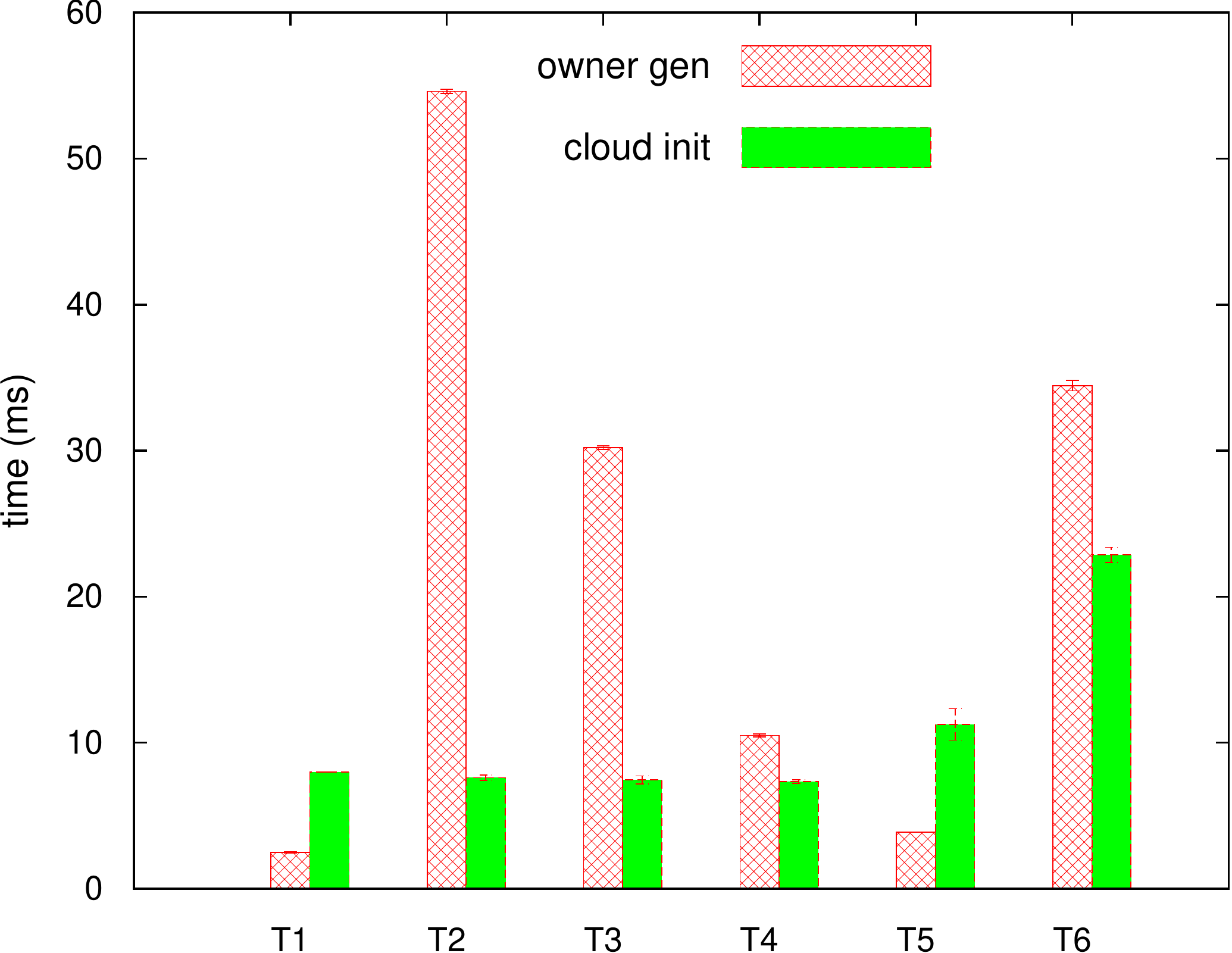}}
\caption{Policy initialization time}
\label{fig:policy}
\end{figure}
The cost to generate and initialize different types of policies are depicted in
Fig.~\ref{fig:policy}. Generating a new policy at the owner involves creating new transformation and
decryption key for the corresponding predicate, which varies with the policy complexity. T2
policies, for example, contain many bag-of-bit attributes that make up complex predicates, and
therefore they take longer. The cost of initializing policies at the cloud depends on key sizes,
hence it is roughly the same for all types of policies, except for Join (which involves 2 keys from the two input
streams). 

\begin{figure}
\hspace*{-1.5cm}
\centering
\subfloat[Transformation]{\includegraphics[scale=0.35]{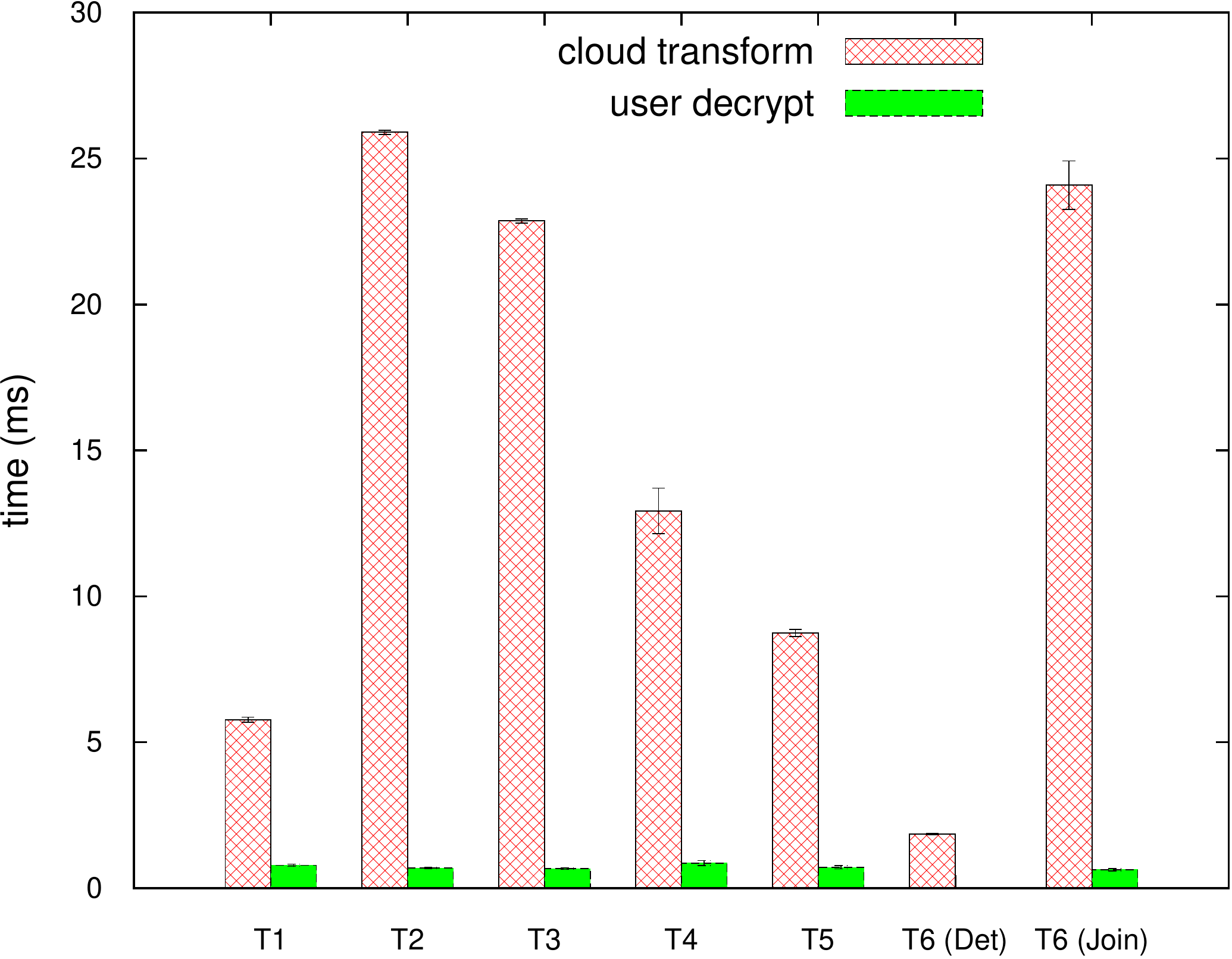}}
\subfloat[Throughput]{\includegraphics[scale=0.35]{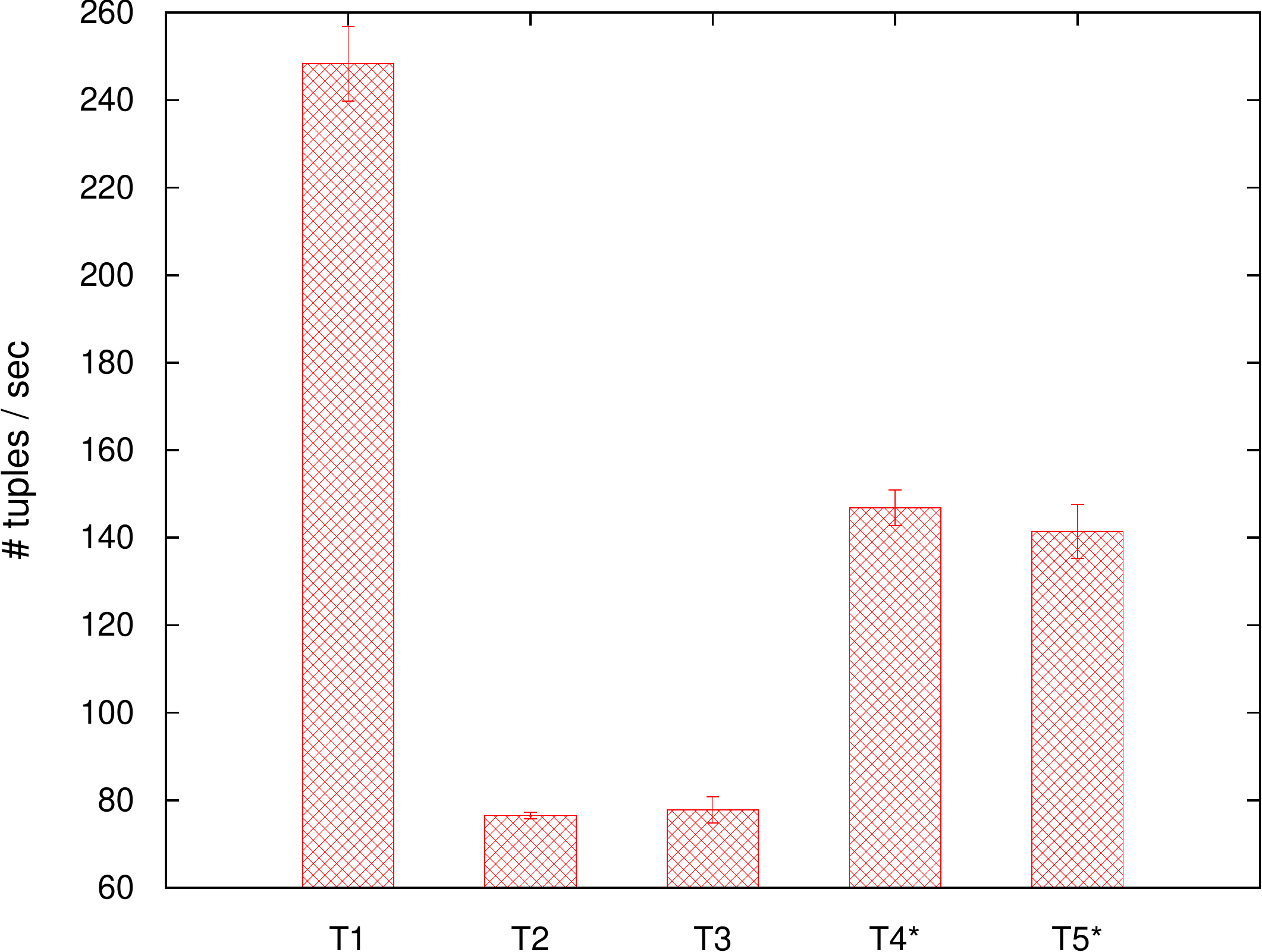}}
\caption{Cost incurred and throughput achieved at the cloud}
\label{fig:cloud}
\end{figure}
Fig.~\ref{fig:cloud}[a] shows the transformation cost at the cloud versus decryption cost at
the user, in which the former is in an order of magnitude bigger. This illustrates that heavy
computations are being outsourced to the cloud. Fig.~\ref{fig:cloud}[b] shows the throughput 
for different policies, in which policies with high transformation cost have low
throughput. The highest throughput is for T1 policies, at $250$ (tuples/sec). Compared to Esper's reported
throughput of over 1 million (tuples/sec), this embarrassingly low figure clearly demonstrates the
high cost of security. However, in many stream applications in practice, such as fitness and weather
monitoring, data arrives at very low rate (in order of minutes). In these cases, our throughput can
sufficiently accommodate many streams at the same time.  Furthermore, as shown later, that
independent streams can be processed in parallel on different servers means the throughput can be
improved by equipping the cloud with more servers.   
      
\begin{figure}
\hspace*{-1.5cm}
\centering
\subfloat[1 stream, multiple policies]{\includegraphics[scale=0.35]{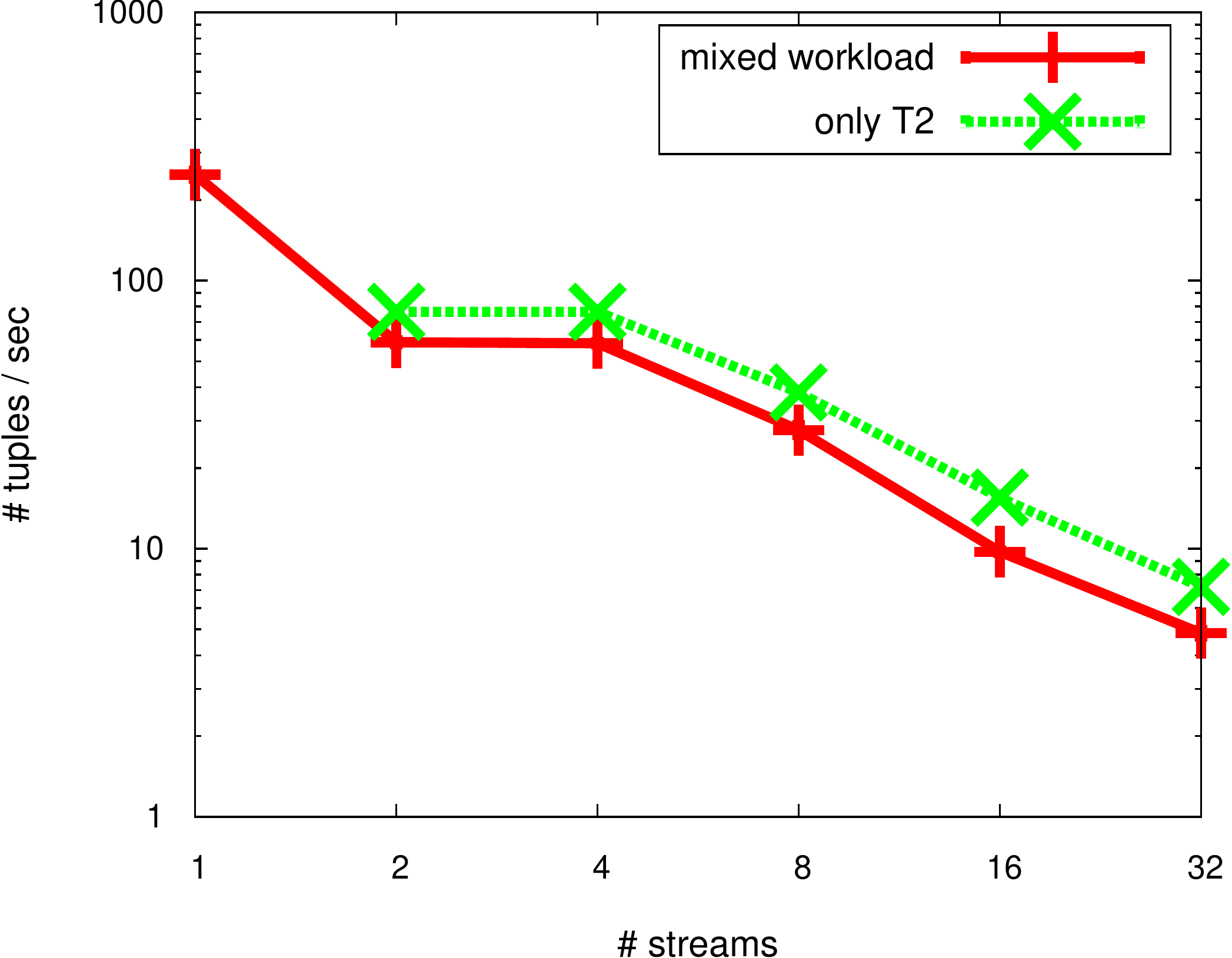}}
\subfloat[multiple streams, multiple policies]{\includegraphics[scale=0.35]{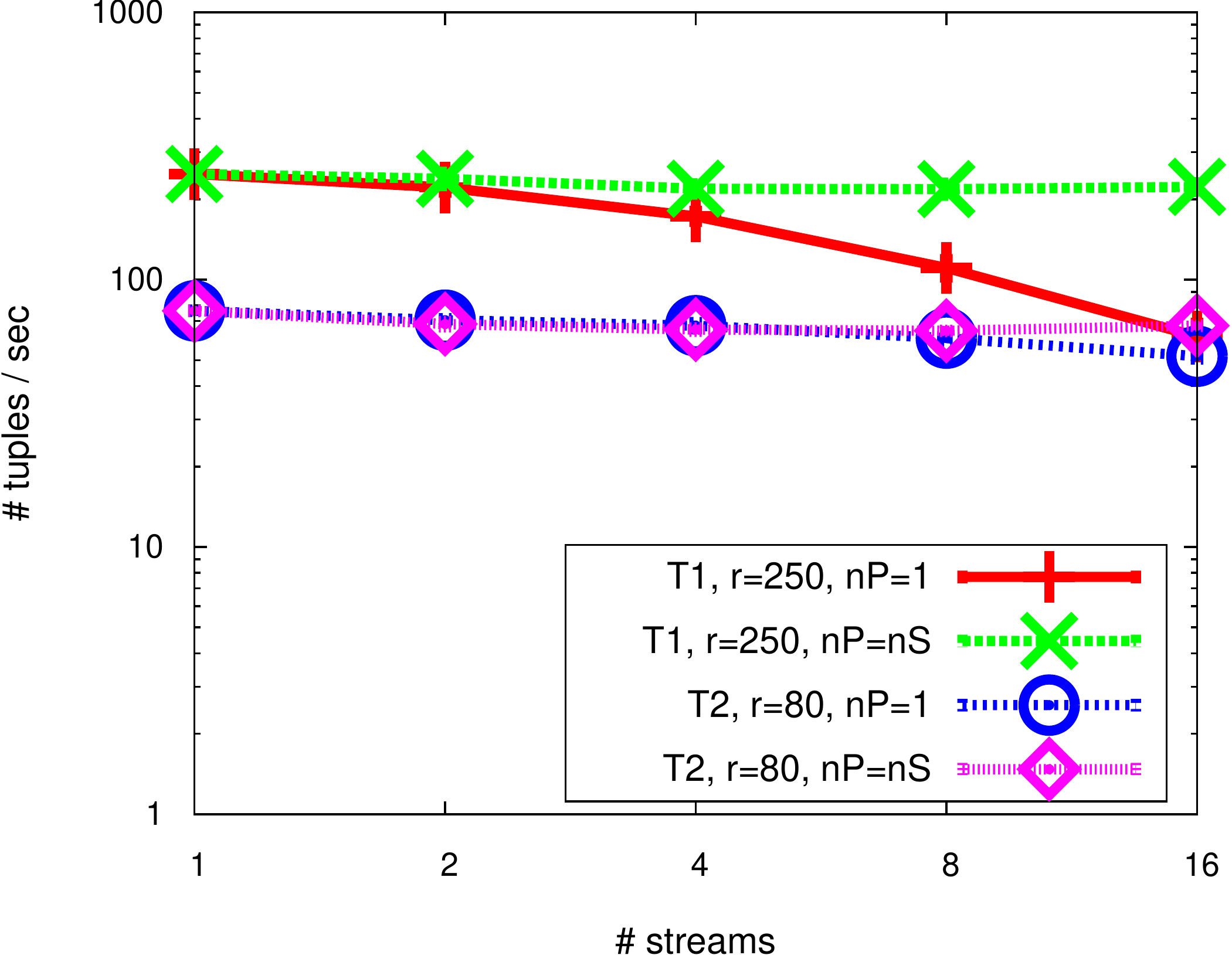}}
\caption{System throughput for complex workloads}
\label{fig:throughput}
\end{figure}

\begin{figure}
\centering
{\includegraphics[scale=0.4]{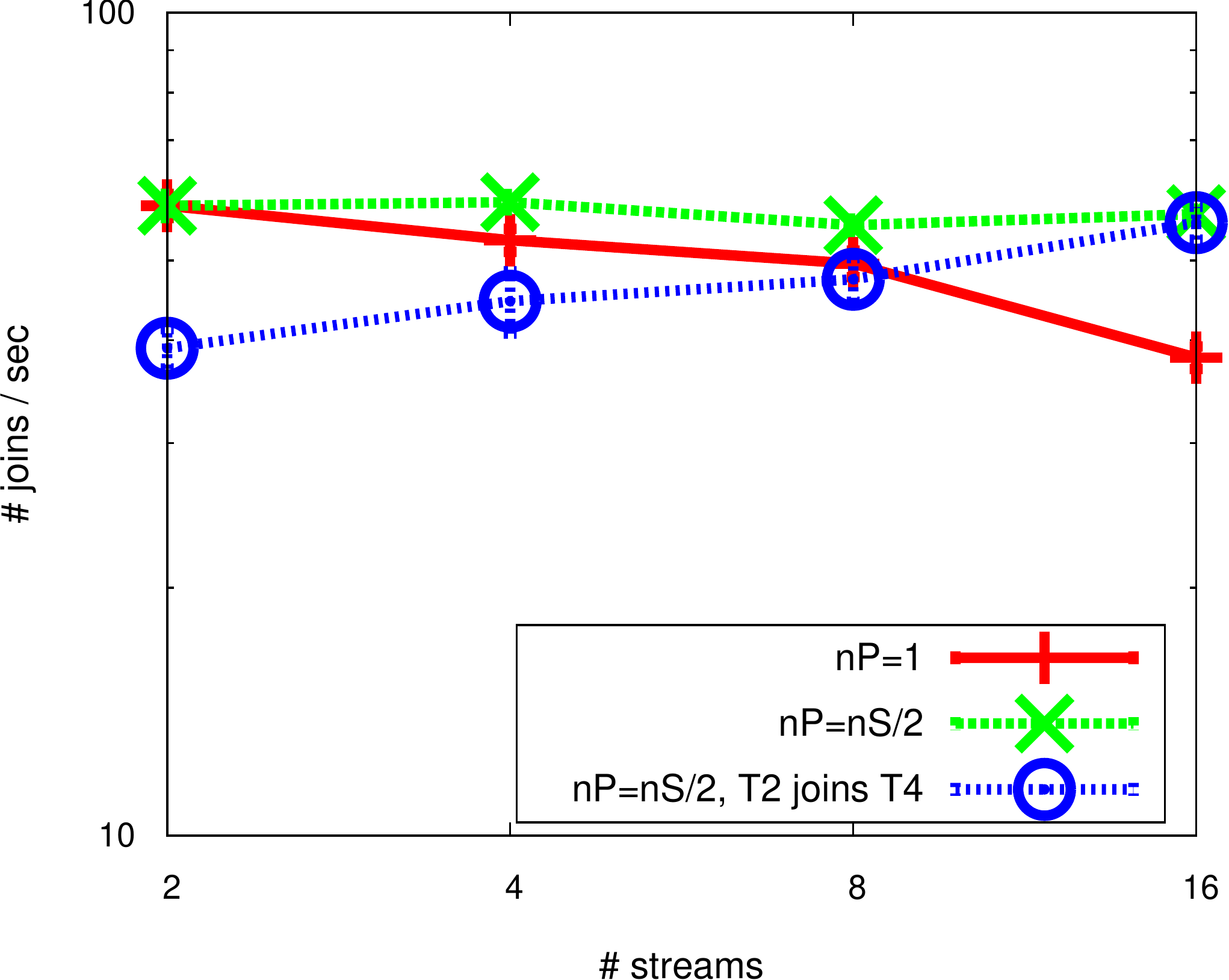}}
\caption{System throughput for join}
\label{fig:throughputJoin}
\end{figure}
Fig.~\ref{fig:throughput} illustrates system throughputs for more complex workloads consisting of multiple
policies, multiple streams. We create mixed workloads containing different types
of policies. Fig.~\ref{fig:throughput}[a] shows that increasing the number of policies decreases the
throughput, which is heavily influenced by the number of T2 policies (the workload of 2 and 4
policies contain only 1 T2 policy). This makes sense because each
tuple has to be matched with (and transformed for) more policies, and because T2's transformation
cost is the highest. When there are multiple streams but only one matching policy, 
communication overheads can reduce the throughput. But as Fig.~\ref{fig:throughput}[b] indicates,
having more matching policies for every stream helps maintain the overall throughput ($r,nP,nS$ are the data rate per
stream, number of matching policies and number of streams respectively). The similar pattern is
found for Join policies, as shown in Fig.~\ref{fig:throughputJoin}. It can be observed that
throughput of join depends on the \emph{similarity} of the two joining
streams.  Specifically, when two Filter conditions are of type T2 $(y < \timestamp < z)$, the output streams (for joining)
have more matches and therefore are more similar (throughput of $60$) than when one filter condition
is of type T4 ($\timestamp \% x = y$) where throughput is at $40$ tuples/sec.   

\begin{figure}
\hspace*{-1.5cm}
\centering
\subfloat[Throughput]{\includegraphics[scale=0.35]{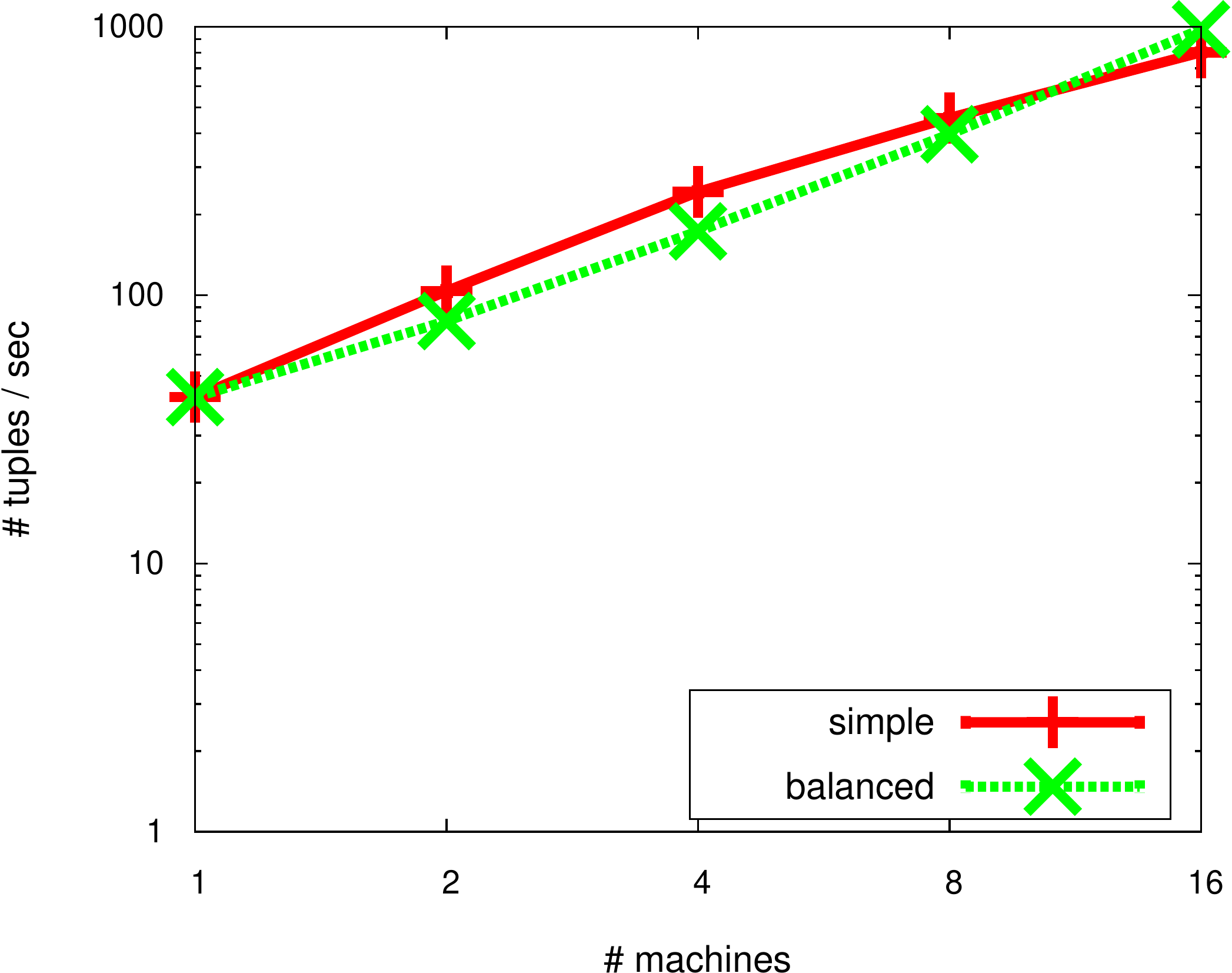}}
\subfloat[Latency]{\includegraphics[scale=0.35]{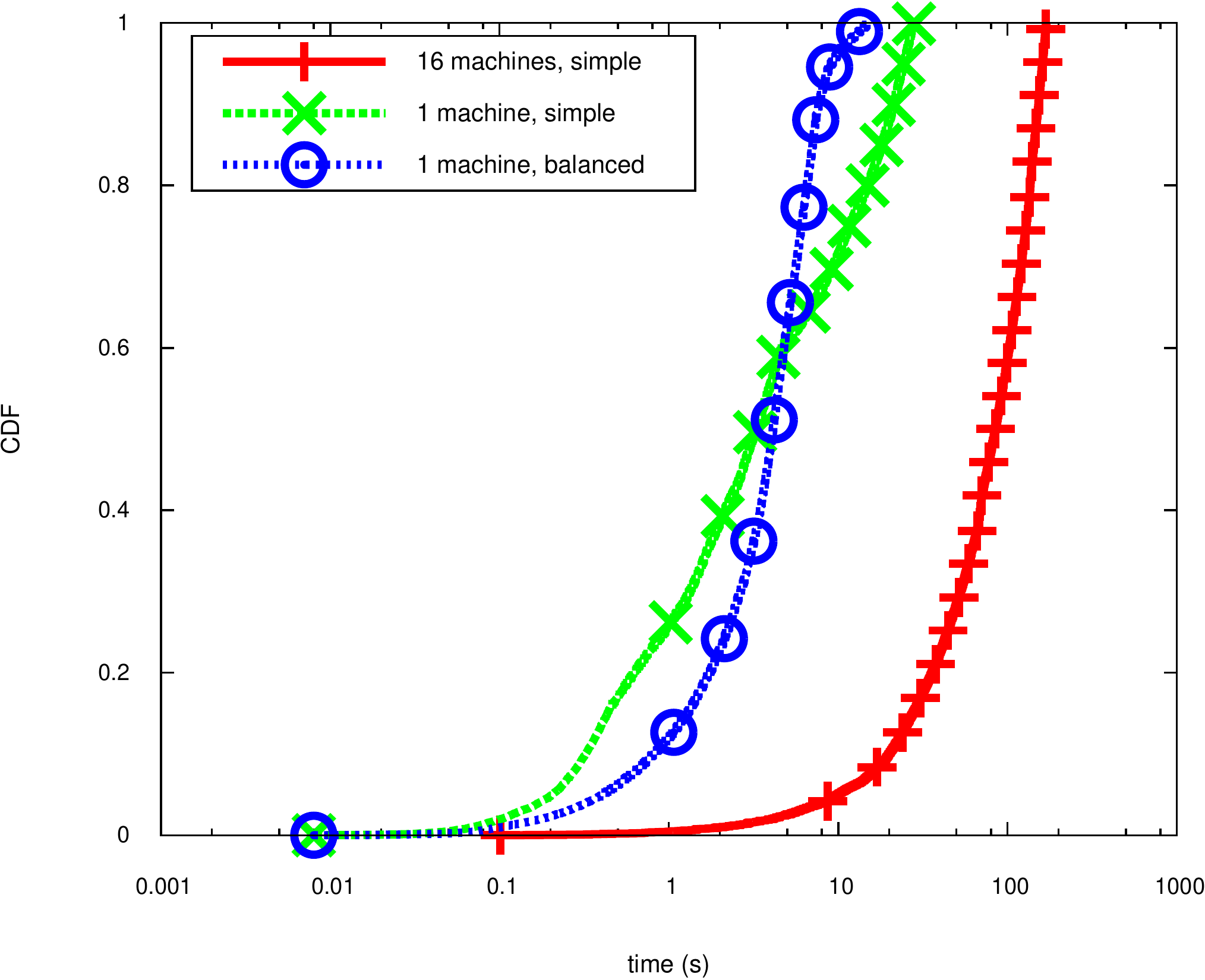}}
\caption{Workload distribution}
\label{fig:scale}
\end{figure}
Finally, Fig.~\ref{fig:scale} illustrates how the system performance improves when more servers
added to the cloud. We create a workload consisting of 16 streams and 320 policies.  4 of these
streams incur expensive load with 4 T2 policies per stream. When there are more than one servers at
the cloud, we consider two ways of distributing the workload: \emph{simple} --- each stream occupies
one machine, and \emph{balanced} --- expensive policies are distributed evenly among the machine.
The latter may result in one stream occupying multiple servers. Fig.~\ref{fig:scale}[a] shows that
the throughput increases linearly with the number of servers, which is as expected. Also, the
balanced distribution achieves lower throughputs, because in the simple distribution the servers
handling light workload gets very high throughputs, whereas with the balanced distribution all servers get low
throughputs. However, at 16 servers, the balanced distribution outgrows the simple distribution, but this
throughput is obtained over duplicate tuples. This is because at 8 and more servers, there are
streams being processed by multiple servers. Fig.~\ref{fig:scale}[b] shows the latency distributions
which clearly demonstrates the benefit of having more servers. The maximum latency using 1 server is
over $100s$, but is reduced to below $14s$ using $16$ machines. The balanced distribution achieves lower maximum latency and lower
variance, since all servers incur a similar load (as opposed to a few servers incurring  much heavier
loads than the others).    

\section{Related Work}
\label{sec:relatedWork}
The design space concerning access control enforcement on a cloud environment can be characterized
using three
properties: policy \emph{fine-grainedness}, cloud \emph{trustworthiness} and cloud/client \emph{work ratio}.
The last property specifies how much work the cloud and user has to perform in relation to each
other: the higher this value, the better it is to move to the cloud. When the cloud is trusted,
it is equivalent to running a private infrastructure, thus the remaining concern is
 policy fine-grainedness. In this setting, ~\cite{carminati10} explores access control model
on top of Aurora query model, while~\cite{dinh12} shows how to extend XACML language to support
fine-grained policies. Such systems achieve the highest level of fine-grainedness. When the cloud is
untrusted, the security must be balanced against the fine-grainedness and work ratio property.
CryptDb~\cite{popa11} ensures data confidentiality against the cloud for archival
database, but it supports only  coarse-grained access control policies. Systems such as~\cite{yu10}
employ ABE schemes for more fine-grained policies, but the work ratio is low because
the cloud only serves as a data storage and distribution facility. 

Streamforce strikes unique balance against all three properties. It considers untrusted cloud (same
as in~\cite{popa11}), supports a wide range of policies (with more fine-grainedness
than~\cite{yu10,popa11}), and at the same time achieves high work ratio (the cloud shares a larger
proportion of the workload than in other systems).

\section{Conclusions and Future Work}
\label{sec:conclusions}
In this paper, we have presented a system providing fine-grained access control for stream data over
untrusted clouds. Our system --- Streamforce --- allows the owners to encrypt data before relaying  
them to the cloud. Encryption ensures both confidentiality against the cloud and access control
against dishonest users. Streamforce uses combinations of three encryption schemes: a deterministic
scheme, a proxy ABE scheme and a sliding-window scheme. We have showed how the cloud can enforce access
control over ciphertexts by transforming them for authorized user, without learning the plaintexts.
In Streamforce, the cloud handles most of the heavy computations, while the users are
required to do only simple, inexpensive decryptions. We have implemented Streamforce on top of Esper, and
carried out a benchmark study of the system. The security cost is large enough to hinder
the system from achieving very high throughputs (as compared to the maximum throughput of Esper).
However, we believe the current throughput is sufficient for many real-life applications in which
data arrives at low rate. Furthermore, we have showed that employing more servers in the cloud can
substantially improve the overall performance. 

We believe that our work has put forth the first secure system for outsourcing the enforcement of
fine-grained access control for stream data. Streamforce occupies an unique position in the design
space, and also opens up a wide avenue for future work. There exists classes of applications that
require much higher throughput than currently possible in Streamforce. We acknowledge that more
effort is required to satisfy both security and demand for performance. However, Streamforce
provides a crucial first step. Our immediate plan is to incorporate our current implementation with
Storm~\cite{lim13}, which deals with workload distribution and helps automate the scaling of our
system. At the current stage, getting the data into Streamforce is the main bottleneck: each
ciphertext is over $100KB$ in size. We are exploring techniques to reduce the ciphertext sizes and
to improve the (incoming) data throughput.

Although Streamforce supports a wide range of policies, this range can still be improved. As
stated in~\cite{dinh13a}, policies involving more complex functions such as granularity and
similarity policies are useful in many applications. Supporting these functions over ciphertext
requires more powerful homomorphic encryptions, such as~\cite{gentry09}. However, one must be careful
to strike the balance between security and performance. Our current encryption schemes do not
support revocation, nor do they support negative and hidden attributes. In particular, hidden
attributes are necessary when the owner wishes to hide more information from the cloud. We plan to
explore if and how existing proposals for these features~\cite{attrapadung08,ostrovsky07,lu13}
can be implemented in our system. Furthermore, we would like to relax the current adversary model
which is semi-honest. A malicious adversary may compromise data integrity, skip computation or
compute using stale data.  We believe that detecting and recovering from these attacks are important
for outsourced database systems, but they may come at heavy cost of performance. Finally, in
Streamforce we have assumed that owners know which data to share and under which policies. In
reality, these decisions are not easy to make. Differential privacy~\cite{dwork06} can be used to
reason about which data to share, while recommendation techniques~\cite{cheek12} can
help determining the appropriate policies.

{\tiny
\bibliographystyle{plain}
\bibliography{paper}
}
\end{document}